\documentclass[
showpacs,
superscriptaddress,
nofootinbib,
floatfix,
11pt]{revtex4-1}
\usepackage{setspace}
\usepackage[utf8]{inputenc}
\usepackage{float}
\usepackage{amsmath,amssymb,amsfonts,bm}
\usepackage{graphicx}
\usepackage{multirow}
\usepackage[usenames,dvipsnames]{color}
\allowdisplaybreaks
\usepackage[
colorlinks=true,
linkcolor=blue,
breaklinks=true,
urlcolor=blue,
citecolor=blue]{hyperref}
\graphicspath{{figs.dir/}}

\definecolor{green}{rgb}{0,0.6,0}

\newcommand{\B}{\color{blue}}
\newcommand{\D}{{\rm d}}
\newcommand{\Tr}{{\rm Tr}}

\newcommand{\be}{\begin{equation}} 
\newcommand{\ee}{\end{equation}}
\newcommand{\bea}{\begin{eqnarray}} 
\newcommand{\eea}{\end{eqnarray}}
\newcommand{\beas}{\begin{eqnarray*}} 
\newcommand{\eeas}{\end{eqnarray*}}

\newcommand{\lag}{\mathcal{L}}
\newcommand{\no}{\nonumber\\}
\newcommand{\tcc}{T_{cc}^+}
\newcommand{\dcn}{D^{*+}D^0}
\newcommand{\dnc}{D^{*0}D^+}

\newcommand{\Br}{{\rm Br}}
\newcommand{\vep}{{\bm p}}
\newcommand{\veq}{{\bm q}}
\renewcommand{\vec}{\bm}

\newcommand{\ific}{\affiliation{Instituto de F\'isica Corpuscular (centro mixto CSIC-UV), \\
Institutos de Investigaci\'on de Paterna, Apartado 22085, 46071, Valencia, Spain}}

\newcommand{\rub}{\affiliation{Institut f\"ur Theoretische Physik II, Ruhr-Universit\"at Bochum, D-44780 Bochum, Germany }}

\newcommand{\fzj}{\affiliation{Institute for Advanced Simulation, Institut f\"ur Kernphysik and J\"ulich Center for Hadron Physics, Forschungszentrum J\"ulich, D-52425 J\"ulich, Germany}}

\newcommand{\itp}{\affiliation{CAS Key Laboratory of Theoretical Physics, Institute of Theoretical Physics, \\Chinese Academy of Sciences, Beijing 100190, China}}

\newcommand{\ucas}{\affiliation{School of Physical Sciences, University of Chinese Academy of Sciences, Beijing 100049, China}}

\newcommand{\itep}{\affiliation{Institute for Theoretical and Experimental Physics NRC ``Kurchatov Institute'', Moscow 117218, Russia }}

\newcommand{\lebedev}{\affiliation{P.N. Lebedev Physical Institute of the Russian Academy of Sciences, 119991, Leninskiy Prospect 53, Moscow, Russia}}

\newcommand{\mipt}{\affiliation{Moscow Institute of Physics and Technology, 141700, Institutsky lane 9, Dolgoprudny, Moscow Region, Russia}}

\newcommand{\scnu}{\affiliation{Guangdong Provincial Key Laboratory of Nuclear Science, Institute of Quantum Matter, South China Normal University, Guangzhou 510006, China}}

\newcommand{\ihep}{\affiliation{Institute of High Energy Physics, Chinese Academy of Sciences, Beijing 100049, China}}

\newcommand{\qmscnu}{\affiliation{Guangdong-Hong Kong Joint Laboratory of Quantum Matter, Southern Nuclear Science Computing Center, South China Normal University, Guangzhou 510006, China}}

\begin{document}
\title{Coupled-channel approach to $T_{cc}^+$ including three-body effects}

\begin{abstract}
A coupled-channel approach is applied to the charged tetraquark state $\tcc$ recently discovered by the LHCb Collaboration. The parameters
of the interaction are fixed by a fit to the observed line shape in the three-body $D^0D^0\pi^+$ channel. Special attention is paid to the three-body dynamics in the $\tcc$ due to the finite life time of the $D^*$. An approach to the $\tcc$ is argued to be self-consistent only if both manifestations of the three-body dynamics, the pion exchange between the $D$ and $D^*$ mesons and the finite $D^*$ width, are taken into account simultaneously to ensure that three-body unitarity is preserved. This is especially important to precisely extract the pole position in the complex energy plane whose imaginary part is very sensitive to the details of the coupled-channel scheme employed. The $D^0D^0$ and $D^0D^+$ invariant mass distributions, predicted based on this analysis, are in good agreement with the LHCb data. The low-energy expansion of the $D^*D$ scattering amplitude is performed and the low-energy constants (the scattering length and effective range) are extracted. The compositeness parameter of the $\tcc$ is found to be close to unity, which implies that the $\tcc$ is a hadronic molecule generated by the interactions in the $D^{*+}D^0$ and $D^{*0}D^+$ channels.
Employing heavy-quark spin symmetry, an isoscalar $D^*D^*$ molecular partner of the $\tcc$ with $J^P=1^+$ is predicted under the assumption that the $ DD^*$-$D^*D^*$ coupled-channel effects can be neglected.

\end{abstract}

\author{Meng-Lin Du}\email{du.menglin@ific.uv.es}
\ific

\author{Vadim Baru}\email{vadim.baru@tp2.rub.de}
\rub \itep 

\author{Xiang-Kun Dong}\email{dongxiangkun@itp.ac.cn}
\itp\ucas 

\author{Arseniy~Filin}
\rub

\author{Feng-Kun~Guo}\email{fkguo@itp.ac.cn}
\itp \ucas 

\author{Christoph~Hanhart}\email{c.hanhart@fz-juelich.de}
\fzj 

\author{Alexey Nefediev}\email{nefediev@lebedev.ru}
\lebedev
\mipt

\author{Juan Nieves }\email{jmnieves@ific.uv.es}
\ific

\author{Qian Wang}\email{qianwang@m.scnu.edu.cn}
\scnu\ihep\qmscnu

\maketitle

\section{Introduction}
\label{sec:intro}

The quest of exotic hadrons with configurations beyond the naive quark-model picture of a pair of quark-antiquark for a meson and three quarks for a baryon has been a central issue in the study of nonperturbative quantum chromodynamics (QCD) for decades. A breakthrough along this path was the discovery of the $X(3872)$ (also known as $\chi_{c1}(3872)$ according to the contemporary classification scheme by the Particle Data Group~\cite{ParticleDataGroup:2020ssz}) by the Belle Collaboration in 2003~\cite{Choi:2003ue}. It resides extremely close to the threshold of a pair of neutral charmed mesons $D^0\bar D^{*0}$. This exotic state is generally considered to be an excellent candidate for a hadronic molecule, which is a composite object formed by at least a pair of hadrons via the strong interaction in analogy to atomic nuclei.
However, since the quantum numbers of the $X(3872)$, $J^{PC}=1^{++}$, are also accessible for a generic $\bar{c}c$ charmonium or a compact tetraquark, debates regarding its internal structure and production mechanisms last since its discovery; see, for example, Refs.~\cite{Gamermann:2009uq,Esposito:2014rxa,Lebed:2016hpi,Chen:2016qju,Guo:2017jvc,Kalashnikova:2018vkv,Yamaguchi:2019vea,Brambilla:2019esw,Guo:2019twa} and references therein for the discussion.

Quite recently, the LHCb Collaboration announced the discovery of a double-charm exotic candidate, $\tcc$, which reveals itself as a high-significance
peaking structure in the $D^0D^0\pi^+$ invariant mass distribution just below the nominal $D^{*+}D^0$ threshold \cite{LHCb:2021vvq}. 
Further studies of the $\tcc$ performed by LHCb \cite{LHCb:2021auc} demonstrate quite intriguing properties of this state and allow for several conclusions concerning its nature: 
\begin{itemize}
\item Narrow near-threshold structures are observed in the $D^0D^0$ and $D^+D^0$ mass spectra, which supports the conjecture that the $\tcc$ decays through a formation of the $D^*$ meson at the intermediate stage of the reaction with its subsequent decays to the $D\pi$ and $D\gamma$ final states,
\beas
&&\tcc\to D^0D^{*+}\to D^0 D^0\pi^+/D^0 D^+\pi^0,\\
&&\tcc\to D^+D^{*0}\to D^+ D^0\pi^0/D^+ D^0\gamma.
\eeas
To produce a visible near-threshold signal in the line shape, the $D^*D$ pair in the $T_{cc}^+$ has to be in $S$-wave. This hints at the quantum numbers of the $\tcc$ to be $J^P=1^+$. 
\item No signal is found in the $D^+D^0\pi^+$ invariant mass distribution, nor is any structure seen in the $D^+D^+$ mass spectrum. This precludes the existence of the $T_{cc}^{++}$ isospin $|I=1,I_3=1\rangle$ state and hints at the $\tcc$ being an isoscalar.
\item The parameters of the resonance extracted from a generic constant-width Breit-Wigner fit built by LHCb in Ref.~\cite{LHCb:2021vvq} are
\beas
\delta m_{\rm BW}&=-273\pm 61\pm 5 ^{+11}_{-14}~\mbox{keV},\\
\Gamma_{\rm BW}&=410\pm 165\pm 43 ^{+18}_{-38}~\mbox{keV},
\eeas
where $\delta m_{\rm BW}$ defines the mass shift, derived from the Breit-Wigner parameterization, with respect to the $D^{*+}D^0$ threshold. However, since approximately 90\% of the $D^0D^0\pi^+$ events contain a genuine $D^{*+}$ meson~\cite{LHCb:2021auc}, it is natural to expect (see the discussions in Refs.~\cite{Feijoo:2021ppq,Yan:2021wdl,Fleming:2021wmk}) that the width of the $\tcc$ should be smaller than that of the $D^{*+}$, which is only $(83.4\pm1.8)$~keV~\cite{ParticleDataGroup:2020ssz}. Thus the value of $\Gamma_{\rm BW}$ quoted above is way too large, suggesting that a more rigorous data analysis is required.
\item A more profound data analysis reported by LHCb in Ref.~\cite{LHCb:2021auc}, based on a unitarized Breit-Wigner parameterization with a momentum-dependent width, allowed to extract the pole position of the amplitude on the second Riemann sheet,
\be
\sqrt{s}_{\rm pole}=\left[-360 \pm 40 ^{+4}_{-0} - i\,(24\pm1 ^{+0}_{-7})\right]~{\rm keV},
\label{LHCbpole}
\ee
where, as before, the real part is given relative to the $D^{*+}D^0$ threshold. The imaginary part of the pole in Eq.~(\ref{LHCbpole}) appears to be in a good qualitative agreement with the natural expectation discussed above. 
\item Only a lower bound was established for a crucial parameter of the model, $g$, which defines the coupling strength of the $\tcc$ to the $D^*D$ channel,
\be
|g|> 5.1(4.3)~\mbox{GeV}~\mbox{at}~90(95)\%~\mbox{CL}.
\ee
The problem is rooted in the intrinsic properties of the resonance, which is quite narrow and located very close to the threshold. In such circumstances, 
the amplitude demonstrates a scaling behaviour~\cite{Baru:2004xg}. In addition, once the employed parameterization is convolved with the energy resolution, the resulting shape appears to be hardly distinguishable from the Breit-Wigner distribution with a constant width. As a consequence, the effective range parameter $r$, which crucially depends on the value of $g$, was extracted to be
\be
0\leqslant -r <11.9\,(16.9)~\mbox{fm}~\mbox{at}~90\,(95)\%~\mbox{CL}.
\label{rLHCb0}
\ee
Similarly, using the formula proposed in Ref.~\cite{Matuschek:2020gqe}, the Weinberg $Z$-factor of the $\tcc$, that is, the probability to find a 
compact component in the $\tcc$ wave function (a component other than $D^{*+}D^0$), was computed using the scattering length and effective range,
\be
Z<0.52\,(0.58)~\mbox{at}~90\,(95)\%~\mbox{CL}.
\label{rLHCb}
\ee
This implies that the properties of the $\tcc$ known so far indicate that this state is 
generally consistent with a molecular nature.
\end{itemize}

Thus from here on we assume that the $\tcc$ is an isoscalar
state and investigate if all of its properties can be
described within a model that treats it as a hadronic molecule.
In the particle basis, the isoscalar character is manifested in the 
equality of the magnitudes of the couplings of the $\tcc$ to the channels $D^{*+}D^0$ and $D^+D^{*0}$, while having opposite signs. However, the $\tcc$ wave function is dominated by the $D^{*+}D^0$ component due to the incredible proximity of the mass of this exotic state to the threshold of this channel.

The discovery of the $\tcc$ quickly spurred a lot of phenomenological studies~\cite{Li:2021zbw,Meng:2021jnw,Agaev:2021vur,Wu:2021kbu,Ling:2021bir,Chen:2021vhg,Dong:2021bvy,Feijoo:2021ppq,Yan:2021wdl,Dai:2021wxi,Weng:2021hje,Xin:2021wcr,Chen:2021kad,Huang:2021urd,Ren:2021dsi,Fleming:2021wmk,Azizi:2021aib,Jin:2021cxj,Hu:2021gdg,Albaladejo:2021vln,Karliner:2021wju}.
However, contrary to the case of the $X(3872)$ with $J^{PC}=1^{++}$, where the static one pion exchange (OPE) interaction is attractive, which allowed T\"ornqvist to correctly predict its mass long before its experimental discovery~\cite{Tornqvist:1993ng}, the static OPE provides 
repulsive and weakly attractive potentials in the isoscalar and isovector $D^*D$ channels with $J^P=1^+$, respectively~\cite{Tornqvist:1993ng,Manohar:1992nd}, 
It should be noted, however, that the static approximation for the OPE in charmonium/double-charm systems of interest here is not justified, since the three-body intermediate state involving the exchanged pion can go on shell. This leads to a potential that has different signs at different values of the momenta. 

An isoscalar bound state in the $D^*D$ system was predicted in quark model~\cite{Janc:2004qn,Carames:2011zz} and in hadronic-level (with the short-distance potential modelled by meson exchanges) calculations~\cite{Yang:2009zzp,Ohkoda:2012hv,Li:2012ss,Liu:2019stu,Liu:2020nil,Ding:2021igr,Dong:2021bvy}. In addition, compact double-charm tetraquarks were also predicted in Refs.~\cite{Ader:1981db,Ballot:1983iv,Zouzou:1986qh,Heller:1986bt,Carlson:1987hh,Silvestre-Brac:1993zem,Silvestre-Brac:1993wyf,Semay:1994ht,Gelman:2002wf,Vijande:2003ki,Navarra:2007yw,Ebert:2007rn,Vijande:2007rf,Zhang:2007mu,Lee:2009rt,Yang:2009zzp,Vijande:2009kj,Abud:2009rk,Wang:2010uf,Valcarce:2010zs,Dias:2011mi,Du:2012wp,Karliner:2013dqa,Feng:2013kea,Luo:2017eub,Karliner:2017qjm,Eichten:2017ffp,Wang:2017uld,Hyodo:2017hue,Cheung:2017tnt,Wang:2017dtg,Richard:2018yrm,Park:2018wjk,Junnarkar:2018twb,Deng:2018kly,Yang:2019itm,Tang:2019nwv,Tan:2020ldi,Lu:2020rog,Braaten:2020nwp,Gao:2020ogo,Cheng:2020wxa,Noh:2021lqs,Faustov:2021hjs}. In particular, it was suggested in Ref.~\cite{Qin:2020zlg} to search for the
$\tcc$ in the channels $D^0D^0\pi^+$ and
$D^0D^+\gamma$ as the LHCb had already collected a
sufficient number of events for the discovery of $\tcc$. For a brief review of the literature, see Ref.~\cite{Dong:2021bvy}. 

In this paper, we present a theoretical analysis of the LHCb data which improves the experimental analysis by LHCb, and the existing theoretical ones in several aspects. 
\begin{itemize}
\item We proceed beyond the simplest approach based solely on the short-range contact interactions (see, for example, the most recent work of Ref.~\cite{Albaladejo:2021vln}) and nonperturbatively include long-range interactions provided by the OPE mechanism. We study its effect on the properties of the $\tcc$ under various assumptions about its form.
\item We study three-body effects in the $\tcc$ state which are expected to have a strong impact on its properties. This is because the $D^{*}$'s are unstable and the corresponding three-body $DD\pi$ thresholds lie very close to and below the two-body $D^*D$ ones, and the $\tcc$ resides between the mentioned two- and three-body thresholds. The interplay of those thresholds in the case of the $X(3872)$
was studied in detail in Ref.~\cite{Baru:2011rs}.
\item We reliably extract the parameters of the effective range expansion from the low-energy scattering amplitude and discuss the consequences for the compositeness of the $\tcc$.
\end{itemize}

Thus, in this work, we investigate the properties of the $\tcc$ in the framework of a nonrelativistic effective field theory constrained with the requirements of isospin and heavy-quark spin (HQSS) symmetries (the leading isospin symmetry breaking is taken into account by using the physical masses of the involved mesons). 

The paper is organized as follows. In Sect.~\ref{sec:framework}, we define our coupled-channel framework including the details of the three-body dynamics. In Sect.~\ref{sec:analysis}, we introduce different fitting schemes and analyse the LHCb data in the $D^0D^0\pi^+$ channel using these schemes. Then we make predictions for the $\tcc$ spin partners in the complementary channels. In Sect.~\ref{sec:effective range expansion}, a low-energy expansion of the scattering amplitude is performed and the low-energy constants (scattering length and effective range) are extracted. An evidence that the $\tcc$ is a composite object is presented in Sect.~\ref{sec:XA}. We discuss the results obtained and conclude 
in Sect.~\ref{sec:disc}.
Generalization to the light flavor SU(3) group is outlined in Appendix~\ref{app:su3} and the effect of a finite width on the effective range is discussed in Appendix~\ref{app:width}. 

\section{Framework}
\label{sec:framework}

\subsection{Interactions}\label{sec:potential}

\subsubsection{Contact potentials}

The leading-order (LO) $D^{(*)}D^{(*)}$ interaction in the chiral effective field theory follows from the effective Lagrangian which contains only $O(p^0)$ contact potentials, with $p$ denoting a small momentum scale~\cite{Mehen:2011yh},
\begin{equation}
\begin{aligned}
\mathcal{L}_{H H} = 
&-\frac{D_{10}}{8}\Tr\left(\tau_{a a^{\prime}}^{A} H_{a^{\prime}}^{\dagger} H_{b} \tau_{b b^{\prime}}^{A} H_{b^{\prime}}^{\dagger} H_{a}\right)
-\frac{D_{11}}{8}\Tr\left(\tau_{a a^{\prime}}^{A} \sigma^{i} H_{a^{\prime}}^{\dagger} H_{b} \tau_{b b^{\prime}}^{A} \sigma^i H_{b^{\prime}}^{\dagger} H_{a}\right)\\
&-\frac{D_{00}}{8}\Tr\left(H_{a}^{\dagger} H_{b} H_{b}^{\dagger} H_{a}\right)
-\frac{D_{01}}{8}\Tr\left(\sigma^{i} H_{a}^{\dagger} H_{b} \sigma^{i} H_{b}^{\dagger} H_{a}\right),
\end{aligned}\label{eq:LHQSS}
\end{equation}
where the subscripts $a^{(\prime)}$, $b^{(\prime)}$ denote flavor indices, $\tau^{A=1,2,3}$ are the isospin Pauli matrices, and 
the $D_{00,10,01,11}$ are four low-energy constants (LECs) describing the contact interactions between the heavy-light mesons grouped into the superfield,
\begin{equation}
H_{a}=P_{a}+\vec V_a \cdot \vec\sigma,
\end{equation}
with $P_a$ and $\vec V_a$ annihilating the ground-state pseudoscalar and vector charmed mesons, respectively, {which in the flavor space are written explicitly as
\begin{equation}
P_a=\left(\begin{array}{c}
D^{ 0} \\
D^{ +}
\end{array}\right)_a, \quad 
\vec V_a=\left(\begin{array}{c}
\vec D^{ *0} \\
\vec D^{ *+}
\end{array}\right)_a.
\end{equation}
}

The proximity of the $\tcc$ to the $D^*D$ thresholds suggests that the dominating component of its wave function consists of a $D$ and $D^*$ meson pair in a relative $S$-wave. The quantum numbers of such a system, $J^P=1^+$, perfectly match the findings of the LHCb Collaboration~\cite{LHCb:2021auc,LHCb:2021vvq}. Then we build the $D^*D$ isoscalar ($I=0$) and isovector ($I=1$) combinations as
\bea
|D^*D,I=0\rangle &=& -\frac{1}{\sqrt{2}}(D^{*+}D^0 - D^{*0}D^+ ),\nonumber\\[-3mm]
\label{eq:iso_decomp}\\[-3mm]
|D^*D,I=1\rangle &=& -\frac{1}{\sqrt{2}}(D^{*+}D^0 + D^{*0}D^+ ),\nonumber
\eea
and employ the Lagrangian of Eq.~\eqref{eq:LHQSS} to find the corresponding $S$-wave contact potentials,
{\bea
V_{\rm CT}^{I=0}(D^*D\to D^*D;1^+)&=&-2(D_{01}-3D_{11})\equiv v_0,\label{CT0}\\
V_{\rm CT}^{I=1}(D^*D\to D^*D;1^+)&=&D_{00}+D_{01}+D_{10}+D_{11}\equiv v_1,\label{CT1}
\eea
where $1^+$ stands for the spin and parity $J^P$.
Then, in the particle basis $\{D^{*+}D^0,D^{*0}D^+\}$, the contact potential reads
\bea
V_{\rm CT}(D^*D\to D^*D; 1^+)=
\begin{pmatrix} 
c&d\\d&c
\end{pmatrix},
\label{candd}
\eea
where the diagonal and off-diagonal matrix elements are
\be
c=\frac12(v_1+v_0),\quad d=\frac12(v_1-v_0).
\label{cd}
\ee
}

According to the claim by LHCb \cite{LHCb:2021auc}, the $\tcc$ is an isoscalar state, so in what follows we stick to the potential of Eq.~\eqref{CT0} and set to zero the contact isovector interaction, that is, $v_1=0$ in Eq.~\eqref{CT1}, or equivalently $d=-c$, to reduce the number of free parameters. The contact potentials in the complementary spin-parity $D^{(*)}D^{(*)}$ channels, as well as their generalization to the light quark flavor SU(3) group can be found in Appendix~\ref{app:su3}. 

In this paper, we work in the strict isospin limit for the contact potentials and take the isospin breaking effects into account through the mass difference of the charged and neutral $D^{(*)}$ mesons as well as that of the pions. 

\subsubsection{OPE potential}

\begin{figure}[t!]
\centering
\includegraphics[width=0.25\linewidth]{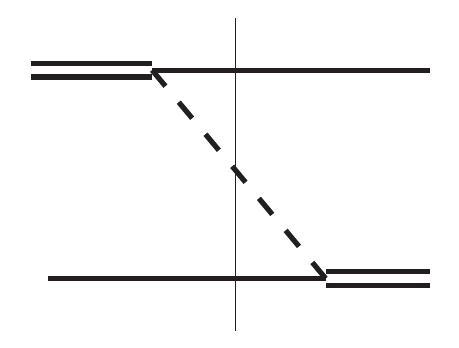}\hspace*{0.15\linewidth}
\includegraphics[width=0.25\linewidth]{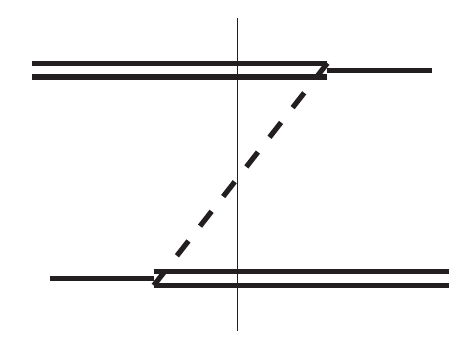}
\caption{The two TOPT contributions ($V_1$ and $V_2$, respectively) to the OPE potential between $D$ (single solid line) and $D^*$ (double solid line) mesons. The dashed line stands for the pion and the vertical thin line shows the relevant intermediate state.}
\label{fig:TOPT}
\end{figure}

The most important portion of the experimental signal is localized within just 1~MeV below the $\dcn$ threshold, while the splitting between the $\dcn$ and $\dnc$ thresholds is around 1.41~MeV, which implies that the isospin breaking effects can be significant. 
We stick to the notations $m_0^{(*)}$ and $m_c^{(*)}$ for the masses of the neutral and charged $D^{(*)}$ mesons, respectively, and
take their values to be~\cite{ParticleDataGroup:2020ssz}
\be
m_0=1864.84~\mbox{MeV},\quad m_c=1869.66~\mbox{MeV},\quad
m_0^*=2006.85~\mbox{MeV},\quad
m_c^*=2010.26~\mbox{MeV}.
\label{mDs}
\ee

The LO Lagrangian for the $D^*D\pi$ interaction reads~\cite{Wise:1992hn,Yan:1992gz,Manohar:1992nd,Fleming:2007rp}
\be
\lag =\frac14g\,\Tr \left({\bm \sigma}\cdot \vec{u}_{ab} H_b H_a^\dag\right),
\label{Lpi}
\ee
where $\vec{u} = -\nabla \Phi/f_\pi$ with 
\be
\Phi=
\begin{pmatrix} \pi^0 & \sqrt{2}\pi^+ \\ 
\sqrt{2}\pi^- & -\pi^0 
\end{pmatrix}.
\ee
Here $f_\pi = 92.1$ MeV is the pion decay constant and the coupling $g=0.57$ is determined from the experimentally measured $D^{*+}\to D^0\pi^+$ decay width. 

The OPE potential can be naturally decomposed into two contributions which correspond to the two different orderings in the framework of the time-ordered perturbation theory (TOPT)---see Fig.~\ref{fig:TOPT}. It should also be noticed that, given the very limited energy and momentum ranges covered by the theory, it is sufficient to employ a nonrelativistic approach for all particles involved, including the pion. Thus, for the propagator of the pion of mass $m_\pi$, we use 
\bea
D^\pi(M, p, p', z) &=& -\frac{1}{2m_\pi}\big[D_1^\pi(M,p, p', z) + D_2^\pi(M,p, p', z) \big], \no
D_1^\pi(M,p, p', z) &=& \left({m_i+m_j+m_\pi+\frac{p^2}{2m_i}+\frac{p^{\prime 2}}{2m_j} + \frac{p^2+p^{\prime 2}-2pp'z}{2m_\pi}-M-i\epsilon}\right)^{-1} ,\label{D1D2}\\
D_2^\pi(M,p, p', z) &=& \left({m_i^*+m_j^*+m_\pi+\frac{p^2}{2m_i^*}+\frac{p^{\prime2}}{2m_j^*} + \frac{p^2+p^{\prime 2}-2pp'z}{2m_\pi}-M-i\epsilon}\right)^{-1}, \nonumber
\eea
where $M$ is the total energy, $\vec{p}$ and $\vec{p}'$ stand for the incoming and outgoing three-momenta, respectively, with $p^{(\prime)}$ their magnitudes, $z=(\vec{p}\cdot\vec{p}')/(p p')$ and $m_{i}^{(*)}$ denotes the mass of the $D^{(*)}$ in the $i$th channel.
To guarantee a proper treatment of the three-body effects, all recoil terms need to be kept in Eq.~(\ref{D1D2}), as discussed in detail in Refs.~\cite{Lensky:2005hb,Baru:2004kw,Baru:2009tx} in the context of the $\pi NN$ and $K NN$ intermediate states. This is especially relevant in the double-charm system at hand given that, near the $\dcn$ threshold, 
$2m_{D^0}+m_{\pi^+}-M\approx m_{D^0}+m_{\pi^+}-m_{D^{*+}}\approx -6~\mbox{MeV},$ and hence the effective parameter which governs the pion exchange (see the first ordering $D_1^\pi$ in Eq.~(\ref{D1D2})), 
\be
\mu_\pi=\sqrt{2 m_\pi (m_{D^0}+m_{\pi^+}-m_{D^{*+}})}\approx i \ 41~\mbox{MeV},
\label{mu4c}
\ee
is not only quite small, but also imaginary. 

The matrix of the OPE potential $V_\text{OPE} (M,p,p')$ in the particle basis, 
\be
\{\dcn(S), \dnc(S), \dcn(D), \dnc(D)\},
\label{basis}
\ee
where $S$ and $D$ in the parentheses indicate the corresponding partial waves, reads
\begin{align}
V_\text{OPE} = \frac{g^2}{12f_\pi^2}
\begin{pmatrix} V_{SS}^{\pi^+} & -\frac12 V_{SS}^{\pi^0} & -{\sqrt{2}}V_{SD}^{\pi^+} & \frac{1}{\sqrt{2}}V_{SD}^{\pi^0} \\
-\frac12 V_{SS}^{\pi^0} & V_{SS}^{\pi^+} & \frac{1}{\sqrt{2}}V_{SD}^{\pi^0} & -{\sqrt{2}}V_{SD}^{\pi^+} \\
-{\sqrt{2}}V_{DS}^{\pi^+} & \frac{1}{\sqrt{2}}V_{DS}^{\pi^0} & \frac{1}{2}V_{DD}^{\pi^+} & -\frac{1}{4}V_{DD}^{\pi^0} \\
 \frac{1}{\sqrt{2}}V_{DS}^{\pi^0} & -{\sqrt{2}}V_{DS}^{\pi^+} & -\frac{1}{4}V_{DD}^{\pi^0} & \frac{1}{2}V_{DD}^{\pi^+} 
\end{pmatrix},
\label{eq:VOPE}
\end{align}
with the individual partial-wave-projected components given by
\bea
V^\pi_{SS}(M,p,p') & = & \int_{-1}^1 \D z \ D^\pi(M,p,p',z) \left( p^2+p^{\prime 2}-2pp'z\right), \no
V^\pi_{SD}(M,p,p') & = & \int_{-1}^1 \D z \ D^\pi(M,p,p',z) \left[\frac12p^2(3z^2-1)+p^{\prime 2}-2pp'z \right],\no[-2mm]
\\[-2mm]
V^\pi_{DS}(M,p,p') & = & \int_{-1}^1 \D z \ D^\pi(M,p,p',z) \left[\frac12p^{\prime 2}(3z^2-1)+p^2-2pp'z \right],\no
V^\pi_{DD}(M,p,p') & = & \int_{-1}^1 \D z \ D^\pi(M,p,p',z) \left[2(p^2+p^{\prime 2})(3z^2-1)-pp'z(9z^2-1) \right].\nonumber
\eea
{It is important to notice that, since the $D^*\to D\pi$ vertex is $P$-wave,  the OPE interaction in the $D^*D$ system at hand contains a short-range contribution and, therefore, is well defined only in the presence of the contact potential introduced above~\cite{Baru:2015nea}. }

\subsection{Lippmann--Schwinger equation}\label{sec:lse}

The dynamics of the system under study can be described in terms of a coupled-channel Lippmann-Schwinger equation (LSE) for the $D^*D\to D^*D$
$T$-matrix (amplitude) satisfying three-body unitarity,
\begin{align}
T_{\alpha\gamma} (M,p, p^\prime)=V_{\alpha\gamma}(M,p, p^\prime) -\sum_\beta \int\frac{\D^3\vec{q}}{(2\pi)^3}V_{\alpha\beta}(M,p,q)G_\beta(M,q)T_{\beta\gamma}( M,q, p^\prime),
\label{eq:lse:scat}
\end{align}
where the Greek indices run from 1 to 4 accounting for the channels listed in Eq.~(\ref{basis}) and
 the potential is treated as a sum of the OPE and contact terms~\cite{Baru:2011rs,Baru:2013rta},
\begin{align}
V(M,p,p')=V_{\rm CT}+V_{\rm OPE}(M,p,p').
\label{Vtot}
\end{align}
The contact potential in the extended basis (\ref{basis}) takes a matrix form,
\be
V_{\rm CT}=\frac{v_0}2
\begin{pmatrix} 
\hphantom{+}1&-1&\hphantom{+}0&\hphantom{+}0\\
-1&\hphantom{+}1&\hphantom{+}0&\hphantom{+}0\\
\hphantom{+}0&\hphantom{+}0&\hphantom{+}0&\hphantom{+}0\\
\hphantom{+}0&\hphantom{+}0&\hphantom{+}0&\hphantom{+}0
\end{pmatrix},
\ee
where, as explained above, we have set to zero the contact isovector interaction and thus taken $c=-d=v_0/2$---see Eqs.~(\ref{candd}) and (\ref{cd}). The 
pion exchange potential $V_{\rm OPE}$ is quoted in Eq.~(\ref{eq:VOPE}).

The full $D D^*$ 
propagators incorporating both the effect of the self-energy from the $D\pi$ loop functions and the contributions from $D\gamma$ decay channels can be expressed as
\begin{align}\label{eq:twoprop}
G_1(M,p) &= G_3(M,p) = \left[{m_{c}^\ast + m_0 + \frac{p^2}{2\mu_{c0}}- M-\frac{i}2\Gamma_c(M,p)}\right]^{-1},\no
G_2(M,p) &= G_4(M,p) = \left[{m_{0}^\ast + m_c + \frac{p^2}{2\mu_{0c}}- M-\frac{i}2\Gamma_0(M,p)}\right]^{-1},
\end{align}
where the reduced masses are $\mu_{c0} = {m_c^\ast m_0}/{(m_c^\ast+m_0)}$ and $\mu_{0c}={m_0^\ast m_c}/{(m_0^\ast+m_c)}$, and the energy-dependent widths read~\cite{Baru:2011rs}
\bea 
\Gamma_c(M,p) & =\Gamma(D^{*+}\to D^+\gamma)&+ \frac{g^2m_0}{12\pi f_\pi^2 m_c^*}\Sigma_{D^0\pi^+D^0}(M,p,\mu_{c0}) +\frac{g^2m_c}{24\pi f_\pi^2 m_c^*} \Sigma_{D^+\pi^0D^0}(M,p,\mu_{c0}),~~ \label{Gamma0}\\
\Gamma_0(M,p) & = \Gamma(D^{*0}\to D^0\gamma) &+ \frac{g^2m_0}{24\pi f_\pi^2 m_0^*}\Sigma_{D^0\pi^0D^+}(M,p,\mu_{0c}) \label{Gammac} \\
& & + \frac{g^2m_c}{12\pi f_\pi^2 m_0^*} \left[\Sigma_{D^+\pi^-D^+}(M,p,\mu_{0c})
-\Sigma_{D^+\pi^-D^+}(m_c+m_0^*,0,\mu_{0c})
\right],\nonumber 
\eea
where 
\be
\Sigma_{ijk}(M,p,\mu) = \left[2\mu_{ij}\left(M-m_i-m_j-m_k-\frac{p^2}{2\mu}\right) \right]^{3/2},\label{Gammaijk}
\ee
with $\mu_{ij}=m_im_j/(m_i+m_j)$.
Notice that $D^{*0}$ has a mass below the $D^+\pi^-$ threshold and the two contributions in the last term in Eq.~(\ref{Gammac}) cancel against each other at the point $p=0$ and $M=m_c+m_0^*$ to ensure that $m_0^*$ represents the physical mass of the $D^{*0}$. The three-body formalism used here incorporates the full three-body dynamics. It is, however, not Lorentz covariant. Meanwhile, since 
the missing diagrams appear only at 
 higher order in the power counting, their omission is justified as shown in Ref.~\cite{Zhang:2021hcl}. For alternative treatments of the three-body dynamics, see, for example, Ref.~\cite{Mikhasenko:2019vhk} and references
therein.

{Since the momentum integrals in the LSE, Eq.~(\ref{eq:lse:scat}), diverge, we regularize them with a sharp cutoff $\Lambda$. The numerical results presented below correspond to $\Lambda=0.5$~GeV. However, we have verified that the physical observables are almost $\Lambda$-independent in a reasonably wide range of $\Lambda$ from 0.3 to 1.2~GeV,
consistent with treating OPE explicitly while effectively integrating out all higher degrees of freedom into contact terms, as given in Eq.~(\ref{Vtot}). A very weak $\Lambda$-dependence of the results obtained should not come as a surprise given a very large separation of scales involved. 
Indeed, the signal in the data is localized within $\Delta\approx 1$~MeV from the $D^*D$ threshold, so that a typical soft scale for the problem at hand can be estimated as $Q\simeq\sqrt{m\Delta}\simeq 40...50$~MeV$\ll\Lambda$, where $m$ is a $D^{(*)}$-meson mass given in Eq.~(\ref{mDs}). Therefore, in the entire interval of the cutoffs used, the EFT expansion parameter $Q/\Lambda$ appears to be extremely small, which makes it possible to absorb the leading-order cutoff dependence in a single momentum-independent contact term. 
Note however that subleading corrections, which scale with the inverse power of the cutoff, still 
contribute to the problem and give rise to some model dependence of the effective range in this leading-order calculation, as discussed at the end of Sec.~\ref{sec:effective range expansion}.  
}

\subsection{Line shape in the $D^0D^0\pi^+$ channel}

\begin{figure}[t!]
\centering
\includegraphics[width=\linewidth]{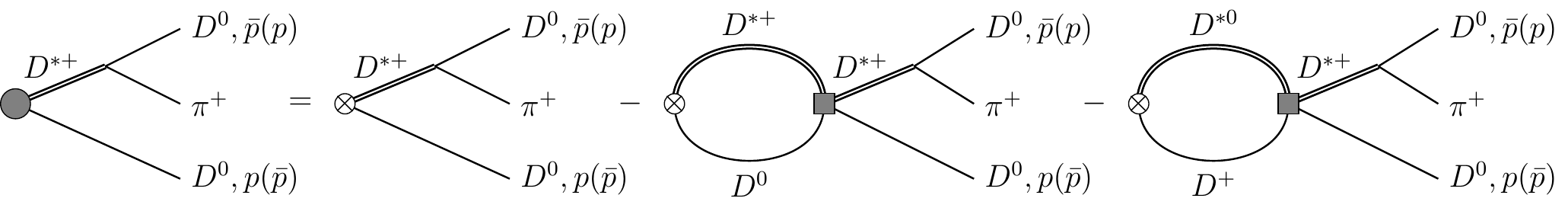}
\caption{Graphical representation for the production amplitude in the $D^0D^0\pi^+$ channel with the $DD^*$ final state interaction. The symbol $\otimes$ stands for the point-like  production source $P_\alpha$, the filled circle is for the full  production amplitude $U_\alpha$ (see Eq.~(\ref{eq:production_J})) while the filled squares stand for the $D^*D$ interactions described by the solution of the LSE quoted in Eq.~\eqref{eq:lse:scat}.}
\label{fig:D0D0pip-feyn}
\end{figure}

{To describe the $D^0D^0\pi^+$ mass distribution, we first proceed from the scattering amplitudes to the production ones, 
so that the 
production amplitude in the $\alpha$th channel, $U_\alpha(M,p)$, takes the form
\be
U_\alpha(M,p) = P_\alpha -\sum_\beta \int\frac{\D^3\vec{q}}{(2\pi)^3} 
T_{\alpha\beta}(M,p,q)G_\beta(M,q)P_\beta, 
\label{eq:production_J}
\ee}
where $P_\alpha$ is a point-like production source for the $\alpha$th channel. In the relatively narrow energy region of interest, we consider only an $S$-wave production. In addition, in the isoscalar channel isospin symmetry requires that $P_2=-P_1$. Finally, since the parameter $P_1$ can always be absorbed by the overall normalization factor, without loss of generality we set $P_1=1$, so that the vector of the sources reads $P_\alpha=(1,-1,0,0)$. 

Then, the production rate for the three-body $D^0D^0\pi^+$ channel (see the corresponding diagrams shown in Fig.~\ref{fig:D0D0pip-feyn}) is calculated as \cite{Baru:2011rs}
\bea\label{eq:lineshape}
\frac{\D\, \Br[D^0D^0\pi^+]}{\D\, M} & =& {\cal N} \int_0^{p_\text{max}}p\,\D p\int_{\bar{p}_\text{min}}^{\bar{p}_\text{max}}\bar{p}\,\D\bar{p}\, \left| q_\pi U_1(M,p) G_1(M,p) +\bar{q}_\pi U_1(M,\bar{p})G_1(M,\bar{p}) \right|^2 , 
\eea
where ${\cal N}$ is a normalization constant,
$$
q_\pi= \sqrt{2\mu_{D^0\pi^+}\left(M-2m_0-m_{\pi^+}-\frac{p^2}{2\mu_p}\right)},\quad 
\bar q_\pi = \sqrt{2\mu_{D^0\pi^+}\left(M-2m_0-m_{\pi^+}-\frac{\bar p^2}{2\mu_p}\right)},
$$
and
$$
p_\text{max} = \sqrt{2\mu_p(M-2m_0-m_{\pi^+})},\quad
\bar{p}_\text{min,max} = \left| \sqrt{2\mu_{D^0\pi^+}\left(M-2m_0-m_{\pi^+}-\frac{p^2}{2\mu_p}\right)} \mp \frac{m_{0}p}{m_{0}+m_{\pi^+}}\right|, 
$$
with $\mu_p=m_0(m_0+m_{\pi^+})/(2m_0+m_{\pi^+})$. 

\subsection{Invariant mass distributions in the $D^0D^0$ and $D^0D^+$ channels}\label{sec:DDspectrum}

\begin{figure}[t!]
\centering
\includegraphics[width=0.45\linewidth]{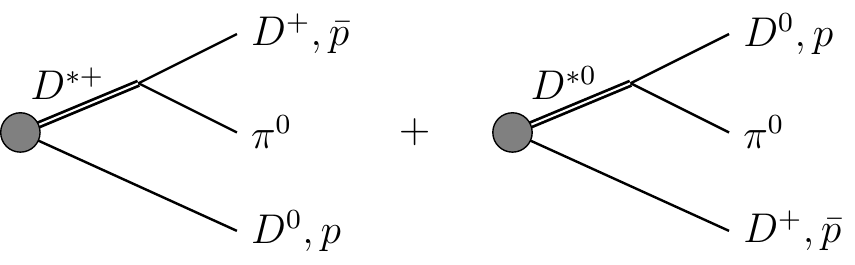}
\caption{Graphical representation for the two contributions to the production amplitude in the $D^0D^+\pi^0$ channel (filled circle) defined in a similar way to that depicted in Fig.~\ref{fig:D0D0pip-feyn}.}
\label{fig:D0Dppi-feyn}
\end{figure}

Invariant mass distributions for the selected $D^0D^0$ (see Fig.~\ref{fig:D0D0pip-feyn}) and $D^0D^+$ (see Fig.~\ref{fig:D0Dppi-feyn}) candidates can be obtained as 
\bea\label{eq:DDspectrum}
\frac{\D\,{\Br[D^0D^0]}}{\D\, m_{00}} = {\cal N}'\int_{m_{00}+m_{\pi^+}}^{M_\text{max}} \D M \int_{m_{23}^\text{min}}^{m_{23}^\text{max}} \D m_{23} \Big| q_\pi U_1(M,p) G_1(M,p) +\bar{q}_\pi U_1(M,\bar{p})G_1(M,\bar{p}) \Big|^2 , \no[-2mm]
\label{Br12}\\[-2mm]
\frac{\D\,{\Br[D^0D^+]}}{\D\, m_{0c}} = {\cal N}'' \int_{m_{0c}+m_{\pi^0}}^{M_\text{max}} \D M \int_{m_{23}^\text{min}}^{m_{23}^\text{max}} \D m_{23} \Big| q_\pi U_1(M,p) G_1(M,p) - \bar{q}_\pi U_2(M,\bar{p})G_2(M,\bar{p}) \Big|^2, \nonumber
\eea
where ${\cal N}'$ and ${\cal N}''$ are normalization constants, $m_{00}$ and $m_{0c}$ are the invariant masses of $D^0D^0$ and $D^0D^+$, respectively, and the particles in the final $D^0D^0\pi^+$ and $D^0D^+\pi^0$ states are labelled as 1, 2 and 3, in order, while the relative negative sign 
in the $D^0D^+$ channel
is due to the isospin relation $g_{D^{*+}\to D^0\pi^+} = \sqrt{2}g_{D^{*0}\to D^0\pi^0} = -\sqrt{2}g_{D^{*+}\to D^+\pi^0}$.\footnote{ In Eq.~\eqref{Br12}, $\D\,{\Br[D^0D^0]}/\D\, m_{00}$ and $\D\,{\Br[D^0D^+]}/\D\, m_{0c}$ should be understood as the number-of-events distributions measured experimentally. Thus, the overall normalization constants depend on the detection efficiency and vary for different final states. The symmetry factor 1/2 in the phase space integration for the $D^0D^0\pi^+$ channel due to the presence of identical $D^0$ mesons is absorbed by ${\cal N}'$. Similarly, the 1/2 factor due to the isospin relation for the axial coupling constants, which appears in the amplitude squared of the $D^0D^+\pi^0$ channel relatively to that of the $D^0D^0\pi^+$ channel, is absorbed by ${\cal N}''$. } Further, in Eq.~(\ref{Br12}),
\bea
q_\pi & = & \sqrt{2\mu_1(M-m_{23}-m_1)}, \qquad \bar q_\pi = \sqrt{2\mu_2(M-m_2-m_{31})}, \no[-2mm]
\\[-2mm]
m_{31} &=& \frac{Mm_{123}-m_{12}(m_1+m_2)-m_{23}(m_2+m_3)+m_1^2+m_2^2+m_3^2}{m_1+m_3},\nonumber
\eea
with $\mu_i = {m_i(m_{123}-m_i)}/{m_{123}}$, $m_{123} = m_1+m_2+m_3$, and $m_{ij}$ the invariant mass of particles $i$ and $j$. The limits of integration $m_{23}^\text{min}$ and $m_{23}^\text{max}$ are determined as
\beas
m_{23}^\text{min} = m_2+m_3+\frac{\mu_{23}\left[ 2 \mu_{12}m_{123}\sqrt{\mu_3(M-m_{12}-m_3)} - m_1m_3\sqrt{2\mu_{12}(m_{12}-m_1-m_2)} \right]^2}{2\mu_{12}^2m_3^2(m_1+m_2)^2},\\ 
m_{23}^\text{max} = m_2+m_3+\frac{\mu_{23}\left[ 2 \mu_{12}m_{123}\sqrt{\mu_3(M-m_{12}-m_3)} + m_1m_3\sqrt{2\mu_{12}(m_{12}-m_1-m_2)} \right]^2}{2\mu_{12}^2m_3^2(m_1+m_2)^2}, 
\eeas
with $\mu_{ij}={m_im_j}/({m_i+m_j})$.

\section{Data analysis}
\label{sec:analysis}

\subsection{Strategy and fitting schemes}\label{sec:fits}

From the consideration of the previous section one can easily see that the $D^*D\pi$ vertex, described by the Lagrangian of Eq.~(\ref{Lpi}), gives rise to two complementary effects: pion exchange in the $D^*D$ system and a momentum-dependent self-energy of the $D^*$, encoded in the momentum-dependent widths given in Eqs.~\eqref{Gamma0} and \eqref{Gammac}. Therefore, a self-consistent treatment of the three-body dynamics requires that both above effects be 
simultaneously included in order not to violate three-body unitarity~\cite{Aaron:1968aoz}. In particular, keeping a nontrivial momentum dependence of the $D^*$ self-energy, thus accounting for the virtual dressing process $D^*\to D\pi\to D^*$, while neglecting the pion exchange between the $D^*$ and $D$ as the $\tcc$ constituents breaks three-body unitarity, which may cause troubles when precisely extracting and interpreting the parameters of the resonance and the low-energy constants in the effective range expansion of the amplitude. 
Thus, in order to assess the role of the three-body effects, we consider the following three different fit schemes:
\begin{itemize}
\item Scheme I (no three-body effects): only the LO contact potentials in the $D^*D$ channels are employed with the constant $D^*$ widths, $\Gamma_0(M,p) = 53.7$ keV and $\Gamma_c(M,p) = 82.5$ keV. This scheme is similar in spirit to the one used in Ref.~\cite{Albaladejo:2021vln}.
\item Scheme II (partial three-body effects): the dynamical widths of the $D^*$ mesons, as given in Eqs.~\eqref{Gamma0} and \eqref{Gammac}, are implemented while the OPE potential is not included.
\item Scheme III (full three-body effects): the complete potential of Eq.~\eqref{Vtot}, which incorporates both the contact and OPE interactions, is used to ensure that the full three-body dynamics is self-consistently taken into account, and in this way three-body unitarity is preserved.
\end{itemize}

We would like to mention that 
a direct comparison of our results with those from the LHCb analysis of Ref.~\cite{LHCb:2021auc} is not possible, since that analysis includes some OPE effects, but not all~\cite{Misha}.

Once the parameters are fixed from the best fit to the data, we search for poles of the amplitude in the complex energy plane. Then the physical $\tcc$ state is associated with the corresponding pole, and its effective coupling $g_\alpha$ to the channel $\alpha$ is obtained from the residue of the scattering amplitude,
\be
g_\alpha g_\beta = \lim_{M\to M_\text{pole}} (M^2 -M_\text{pole}^2) T_{\alpha\beta}(M).
\label{gab}
\ee
Here the on-shell scattering amplitude $T_{\alpha\beta}(M)$ is evaluated from Eq.~\eqref{eq:lse:scat} by imposing the condition that $p^{(\prime)}$ is the pole momentum of the two-body propagator $G_{\alpha(\beta)}(M,p)$, which 
reduces to Eq.~\eqref{eq:erepoint} given below for a constant $D^*$ width.

\subsection{Fits to the $D^0D^0\pi^+$ spectrum and predictions for the $D^0D^0$ and $D^0D^+$ line shapes}\label{sec:DDpispectrum}

\begin{figure*}[t!]
\centering
\includegraphics[width=0.5\textwidth]{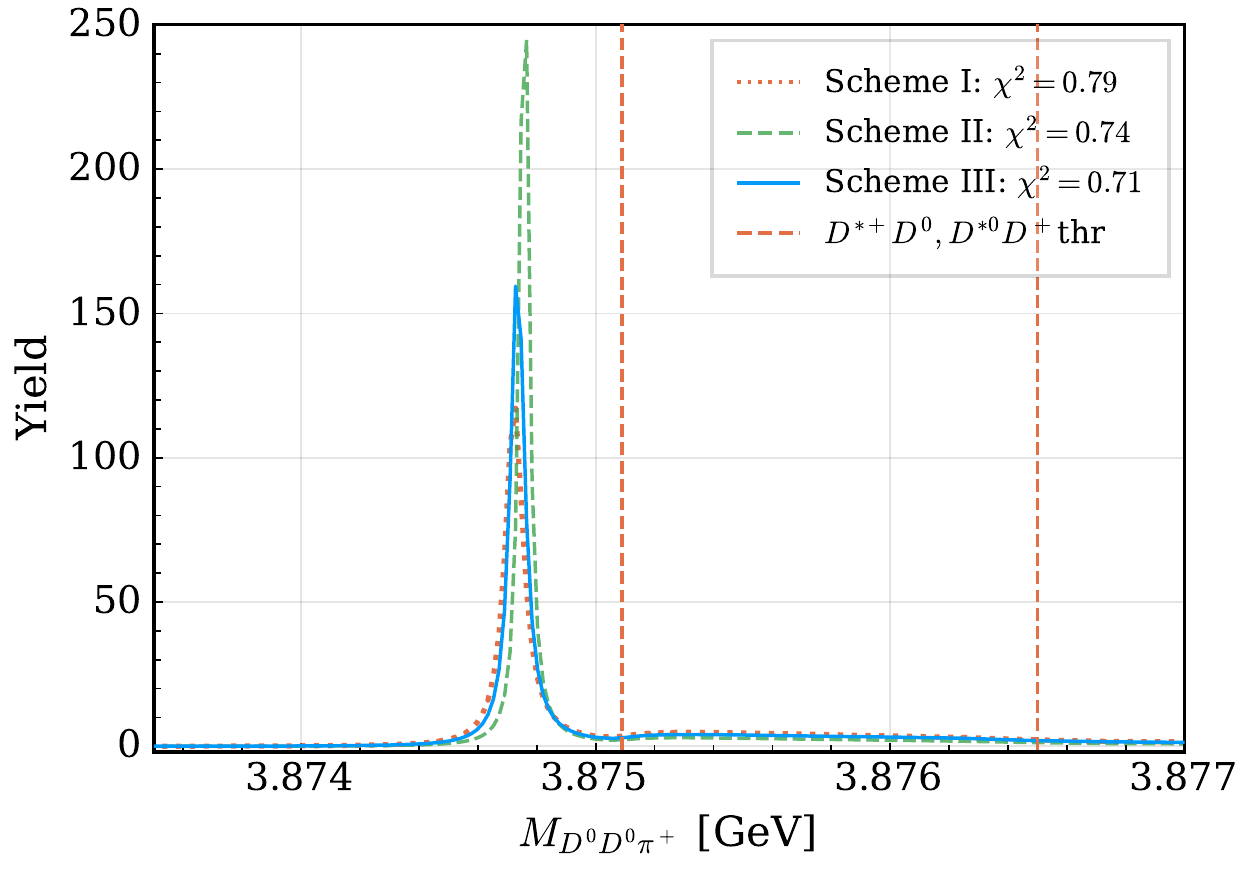}\includegraphics[width=0.5\textwidth]{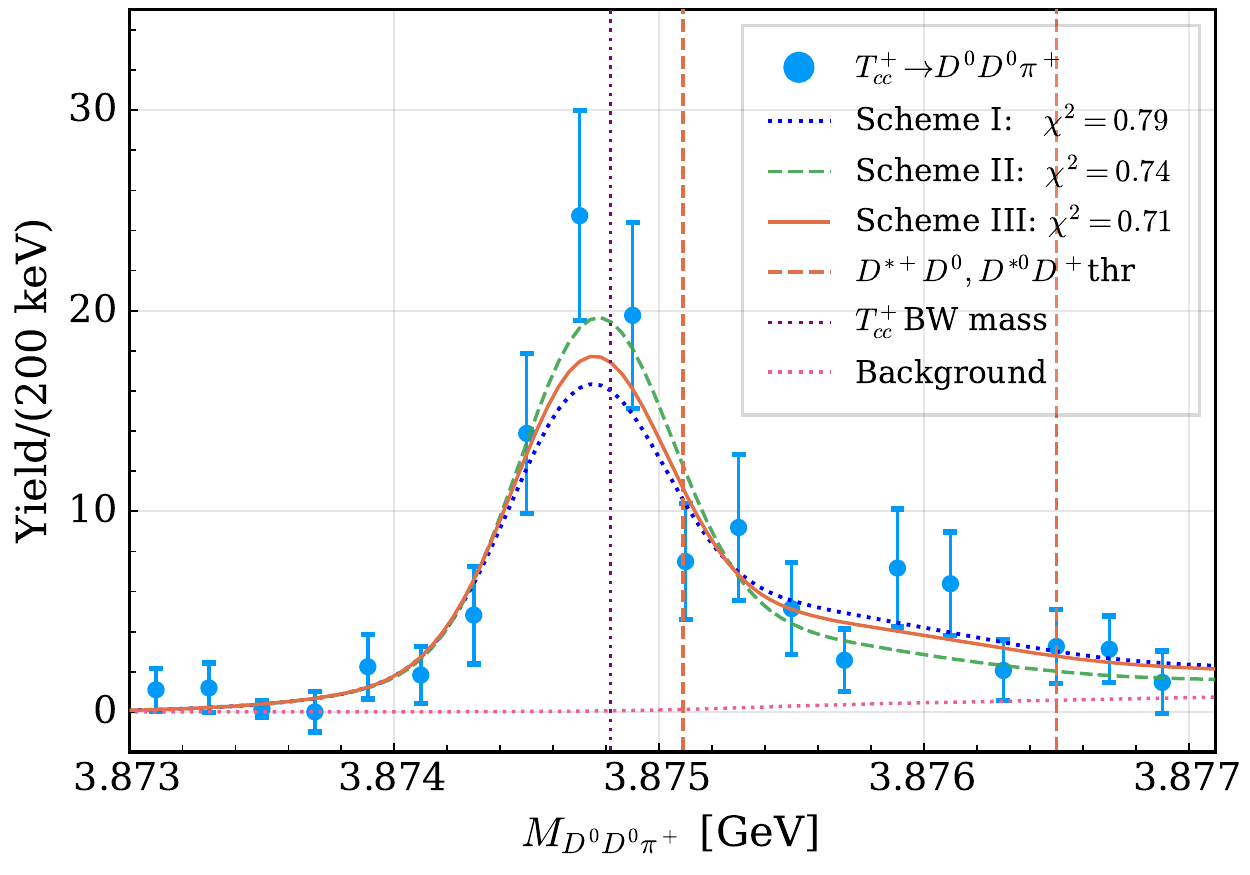}
\caption{Fitted line shapes before (left plot) and after (right plot) convolution with the energy resolution function---see footnote~\ref{resfun} for its explicit form. The background is taken from the LHCb analysis~\cite{LHCb:2021vvq,LHCb:2021auc}.
The experimental binning with the bin size of 200 keV is included in the fits.} 
\label{fig:fits}
\end{figure*}

\begin{table*}[t!]
\begin{ruledtabular}
\caption{Values of $\chi^2/{\rm d.o.f.}$ for the three fit schemes, together with the fitted values of $v_0$ (see Eq.~(\ref{CT0})). The cutoff in the LSE is set to $\Lambda=0.5$~GeV.}
\begin{tabular}{lccc}
Scheme& I & II & III \\
\hline
$\chi^2/{\rm d.o.f.}$ & 0.79 & 0.74 & 0.71 \\
\hline
$v_0$ [GeV$^{-2}]$ & $-23.34\pm0.08 $ & $-22.88^{+0.08}_{-0.06}$ & $-5.04^{+0.10}_{-0.08}$ 
\end{tabular}
\label{tab:fits}
\end{ruledtabular}
\end{table*}

The expression of Eq.~\eqref{eq:lineshape} for the line shape in the channel $D^0D^0\pi^+$ contains only two free parameters: the strength of the contact potential $v_0$ from Eq.~(\ref{CT0}) and the overall normalization factor ${\cal N}$, with the shape depending only on the former one. The signal function is supplemented with the combinatorial background taken directly from the LHCb analysis of Refs.~\cite{LHCb:2021vvq,LHCb:2021auc}. In addition, the experimental energy resolution is taken into account using the resolution function given in Ref.~\cite{LHCb:2021auc}.\footnote{\label{resfun}The resolution function for the $\tcc$ mass distribution, $\mathfrak{R}_\text{LHCb}$, is parameterized by a sum of two Gaussian functions, 
\bea\label{Eq:resol}
\mathfrak{R}_{\rm LHCb}(M, M^\prime) = \alpha \mathcal{G}(M,M^\prime,\sigma_1) + (1-\alpha) \mathcal{G}(M,M^\prime,\sigma_2),
\quad \mathcal G(x,\mu, \sigma)=\dfrac{1}{\sqrt{2\pi}\sigma} \exp\left(-\frac{(x-\mu)^2}{2\sigma^2}\right),
\eea
with the parameters $\sigma_1 = 1.05\times 263$ keV, $\sigma_2=2.413\times \sigma_1$, and $\alpha = 0.778$ taken from Ref.~\cite{LHCb:2021auc}. }
The fit results for the three schemes introduced in Sect.~\ref{sec:fits} are shown in Fig.~\ref{fig:fits}, and the best fit parameters are collected in Table~\ref{tab:fits}. The quality of each fit can be assessed through the corresponding value of $\chi^2/{\rm d.o.f.}$ quoted in the same table. The pole position responsible for the $\tcc$ in each scheme is given in Table~\ref{tab:poles}. The real and imaginary parts of the pole position are treated as the binding energy and half of the width, respectively. 

In Scheme I, where a constant width of the $D^*$ is employed and hence the three-body cut does not show up, the Riemann surface contains four sheets. Then the $\tcc$ pole is located on the first (physical) Riemann sheet (RS-I), just below the $\dcn$ threshold and thus it describes a shallow bound state. 
This Riemann sheet (RS) corresponds to positive values of the imaginary part of all involved momenta. It has to be noticed, however, that by including a constant $D^*$ width (or, equivalently, a complex $D^*$ mass), one distorts the two-body cut, which does not any longer spread along the real axis, so that the bound state pole below the threshold on RS-I naturally acquires an imaginary part. 

In Schemes II and III, when the three-body channels ($DD\pi$) are included explicitly, the three-body cuts appear with their branch points at the 
three-body thresholds. Thus, in these two schemes, the $\tcc$ pole is located in the lower half plane of the second Riemann sheet (RS-II)\footnote{Note that here RS-II is not the second Riemann sheet in the usual sense in two-body scattering. Instead, we use it to refer to the unphysical RS, specified by Eq.~\eqref{eq:continuation}, with respect to the branch points at the three-body thresholds. } and its position can be accessed through the analytic continuation of the self-energy~\cite{Doring:2009yv},
\be
\Sigma_{ijk}(M,p,\mu) \to \begin{cases}
-\Sigma_{ijk}(M,p,\mu), & \mathfrak{Im} M < 0\ \&\ \mathfrak{Re}\left(M-m_i-m_j-m_k-\frac{p^2}{2\mu}\right) > 0, \\
\Sigma_{ijk}(M,p,\mu), 
& \mathfrak{Im} M < 0\ \&\ \mathfrak{Re}\left(M-m_i-m_j-m_k-\frac{p^2}{2\mu}\right) < 0, 
\end{cases} 
\label{eq:continuation}
\ee
with $\Sigma_{ijk}$ defined in Eq.~(\ref{Gammaijk}). 
Then, in the energy range near the $\tcc$ pole, 
the $D^0D^0\pi^+$ and $D^+D^0\pi^0$ channels are on their unphysical RSs while the $D^+D^+\pi^-$ is on its physical RS.
 
\begin{table*}[!t]
\small
\begin{ruledtabular}
\caption{The pole position of the $\tcc$ relative to the $\dcn$ threshold and the Riemann sheet (RS) where the pole is located in each scheme (see the text for details). The  errors are statistical propagated from fitting to the LHCb data while the uncertainties from the cutoff variation are well within the errors quoted here.
}
\begin{tabular}{l|ccc}
Scheme& I & II & III \\
\hline
Pole [keV] & $-368^{+43}_{-42} - i(37\pm0)$ (RS-I) & $-333^{+41}_{-36}-i(18\pm1)$ (RS-II) & $-356^{+39}_{-38}-i(28\pm1)$ (RS-II) 
\end{tabular}
\label{tab:poles}
\end{ruledtabular}
\end{table*}

\begin{table*}[!t]
\small
\begin{ruledtabular}
\caption{Effective couplings extracted as indicated in Eq.~(\ref{gab}). Note that in Schemes II and III, the couplings are complex, with non-zero imaginary parts, although much smaller than the corresponding real parts. The dots mean that the D wave coupling is not included.
}
\begin{tabular}{l|ccc}
Scheme& I & II & III \\
\hline
$g_{\dcn}(S)$ & $1.03\pm0.03$ & $(1.00\pm0.03)-i(0.01\pm0.00)$ & $(1.03\pm0.02)-i(0.01\pm0.01)$ \\
\hline
$g_{\dnc}(S)$ & $-1.03\pm0.03$ & $(-1.00\pm0.03)+i(0.01\pm0.00)$ & $(-0.99\pm0.02)+i(0.01\pm0.01)$ \\
\hline
\hline
$g_{D^*D}^{(I=0)}(S)$ & $-1.45\pm0.04$ & $(-1.42\pm0.03)+i(0.01\pm0.00)$ & $(-1.43\pm0.03)+i(0.02\pm0.00)$ \\
\hline
$g_{D^*D}^{(I=1)}(S)$ & $0.00\pm0.00$ & $(0.00\pm0.00)+i(0.00\pm0.00)$ & $(-0.03\pm0.00)+i(0.00\pm0.00)$\\
\hline
$g_{D^*D}^{(I=0)}(D)$ & $\cdots$ & $\cdots$ & $(0.02\pm0.00)+i(0.00\pm0.00)$ \\
\hline
$g_{D^*D}^{(I=1)}(D)$ & $\cdots$ & $\cdots$ & $(-0.00\pm0.00)+i(0.00\pm0.00)$ 
\end{tabular}
\label{tab:couplings}
\end{ruledtabular}
\end{table*}

A comment on the role played by the three-body dynamics in the $\tcc$ is in order here. As mentioned above, the imaginary part of the pole can be treated as half of the $\tcc$ width. It is, therefore, instructive to notice that neglecting the three-body dynamics due to the finite life time of the $D^*$ one overestimates the $\tcc$ width by up to a factor of 2 (compare the imaginary parts of the pole positions for Schemes I and II quoted in Table~\ref{tab:poles}). 
This shift is partially overcome once also the three-body cut is included in the scattering potential. 
If after neglecting the three-body effects, one employs the static approximation for the OPE, together with a constant width of the $D^*$, the half width of the $\tcc$ would turn out to be around 70 keV thus overestimating the full result in Scheme III by a factor of 2.5.
These results agree with the conclusions obtained in Ref.~\cite{Baru:2011rs} about the role of the three-body $D\bar{D}\pi$ dynamics in the $X(3872)$.

In Table~\ref{tab:couplings}, we compile the values of the effective coupling constants to the different channels extracted as detailed in Eq.~(\ref{gab}). 
For convenience, we also define these couplings in the isospin basis.
In Schemes I and II, where only isospin conserving contact interactions are retained, the coupling of the $\tcc$ to the isovector channel vanishes exactly, while in Scheme III, because of the isospin symmetry breaking effects in the OPE driven by the mass difference between the charged and neutral pions, there is a non-vanishing, though very small, admixture of the isovector component.
The $D$-wave couplings in Scheme III (given only in the isospin basis) are strongly suppressed compared with the $S$-wave ones. This should not come as a surprise given the very limited near-threshold energy range spanned by the signal. 

With the contact interaction parameter $v_0$ (see Eq.~(\ref{CT0})) determined from the fit, it is straightforward to predict the invariant mass distributions in the $D^0D^0$ and $D^0D^+$ channels as given in Eq.~\eqref{eq:DDspectrum}, and the results are presented in Figs.~\ref{fig:D0D0} and \ref{fig:D0Dc}. In both figures, the left and right panels show the corresponding distributions before and after convolution with the experimental energy resolution function, respectively. For the latter, we use the same form as that for the experimental $D^0D^0\pi^+$ mass distribution with the two Gaussian functions given in Eq.~\eqref{Eq:resol}---see footnote~\ref{resfun}. 
Both distributions present a narrow peak just above the corresponding $DD$ threshold, which is a reflection of the $\tcc$, as a very shallow bound state of the $D$ and $D^*$, decaying into $DD\pi$ through the intermediate off-shell $D^*$ with a very small energy released in 
the transition $D^*\to D\pi$. Thus the form of the $DD$ mass distributions supports the interpretation of the $\tcc$ as a weakly bound $D^*D$ molecule. For a calculation of the same distributions in a
formalism with perturbative pions, we refer to the work in Ref.~\cite{Fleming:2021wmk}.
A more precise measurement of the $DD$ invariant mass distributions would be helpful to understand the $DD$ interaction.\footnote{The impact of $D\bar D$ interactions on the $X(3872)\to D^0\bar D^0\pi^0$ decay is studied in Refs.~\cite{Guo:2014hqa,Dai:2019hrf}. It is found in Ref.~\cite{Guo:2014hqa} that the impact can be sizeable if there is a near-threshold $D\bar D$ bound/virtual state. Analogously, from the nice agreement of the $DD$ invariant mass distributions in Figs.~\ref{fig:D0D0} and \ref{fig:D0Dc}, one is tempted to conclude that there is likely no near-threshold $DD$ bound/virtual state, consistent with the spectrum predicted in Ref.~\cite{Dong:2021bvy}. }

\begin{figure*}[t!]
\centering
\includegraphics[width=0.5\textwidth]{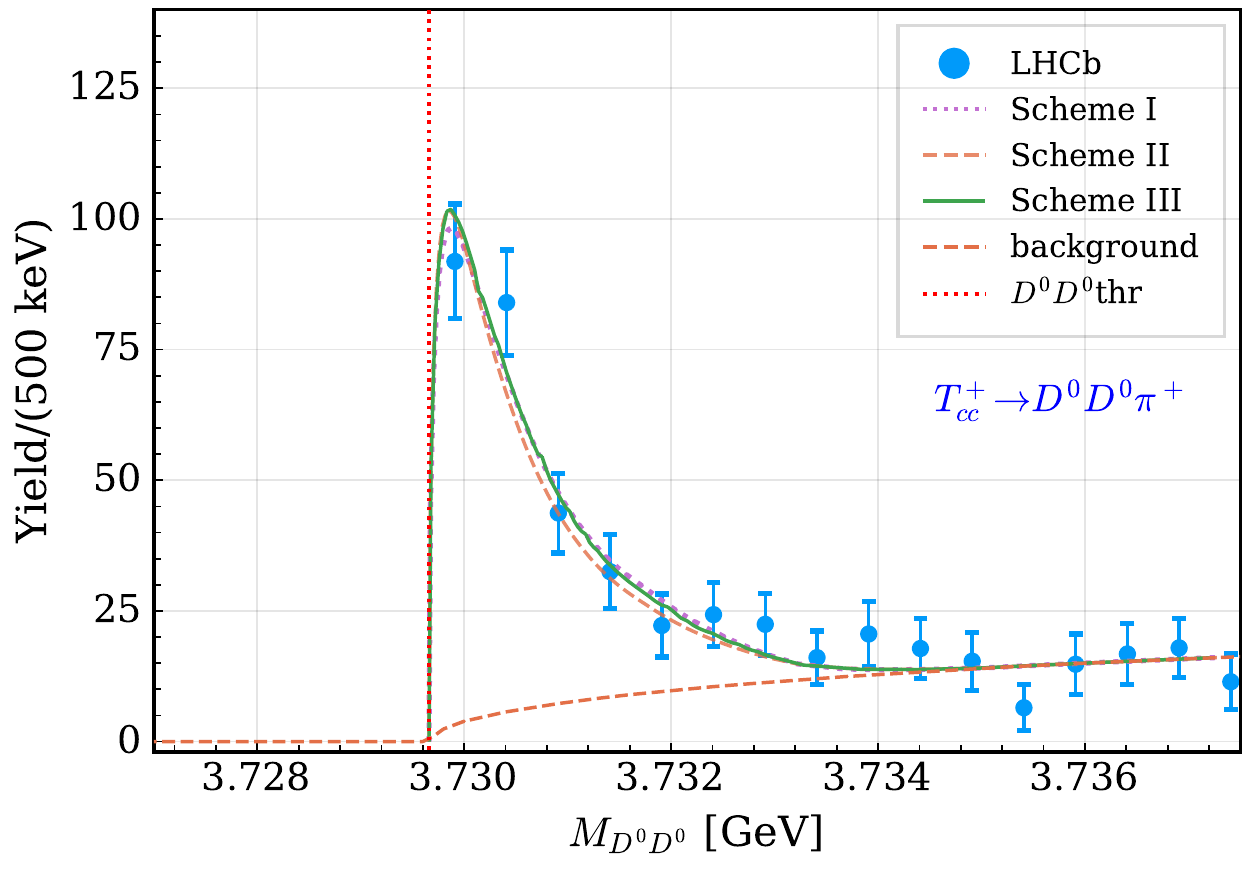}\includegraphics[width=0.5\textwidth]{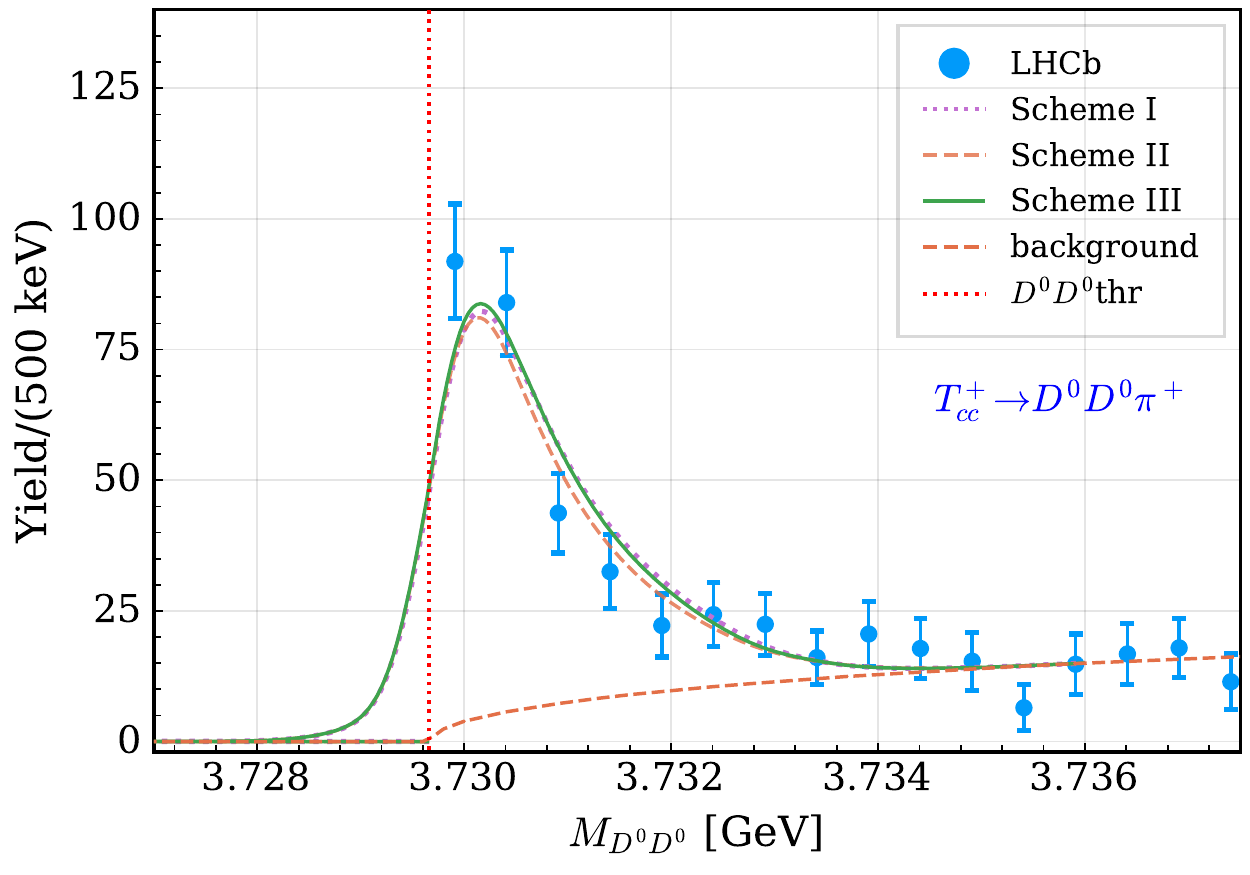}
\caption{The $D^0D^0$ invariant mass distribution of the $T_{cc}^+\to D^0D^0\pi^+$ decay predicted using the contact interaction parameter $v_0$ (see Eq.~\eqref{CT0}) determined from fitting the $D^0D^0\pi^+$ spectrum (see Table~\ref{tab:fits}) before (left plot) and after (right plot) convolution with the energy resolution function.}
\label{fig:D0D0}
\end{figure*}

\begin{figure*}[t!]
 \centering
\includegraphics[width=0.5\textwidth]{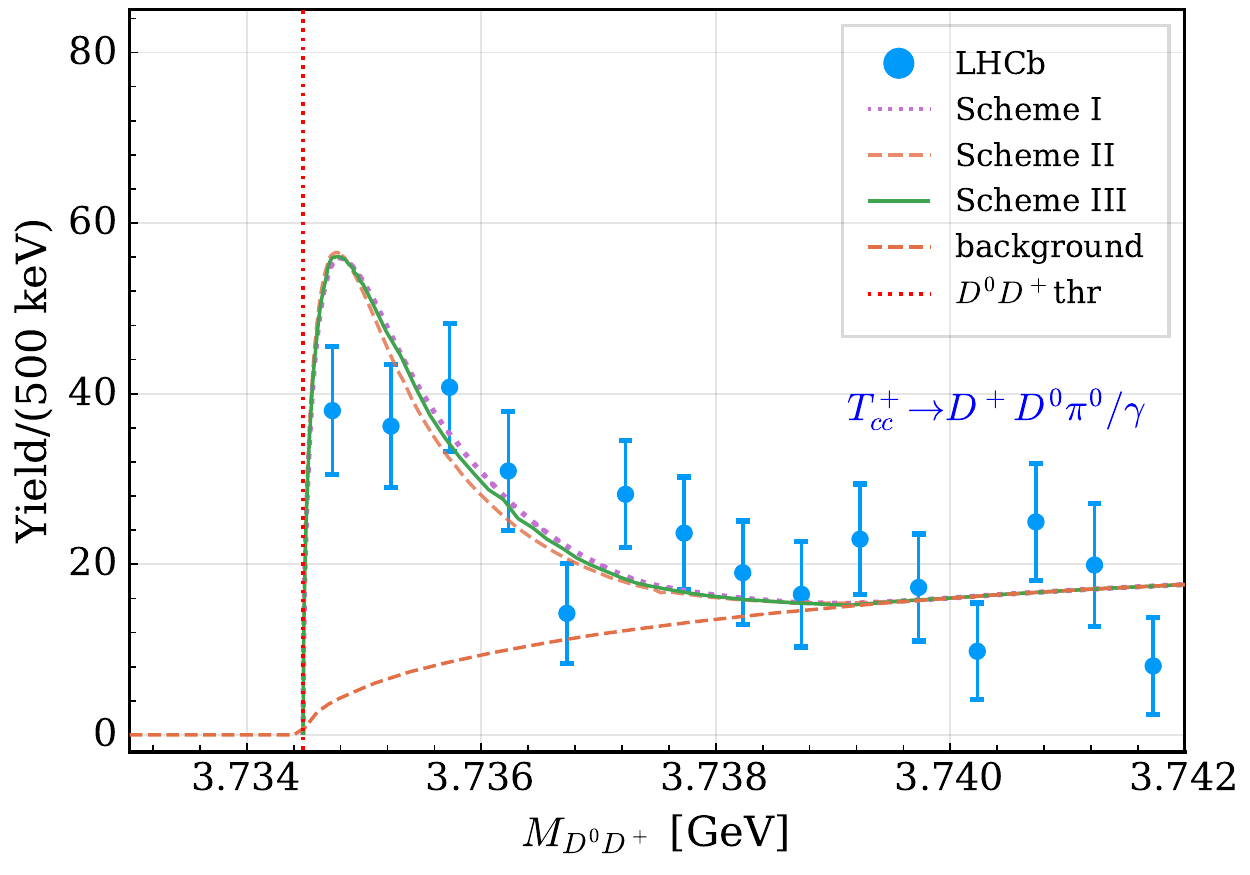}\includegraphics[width=0.5\textwidth]{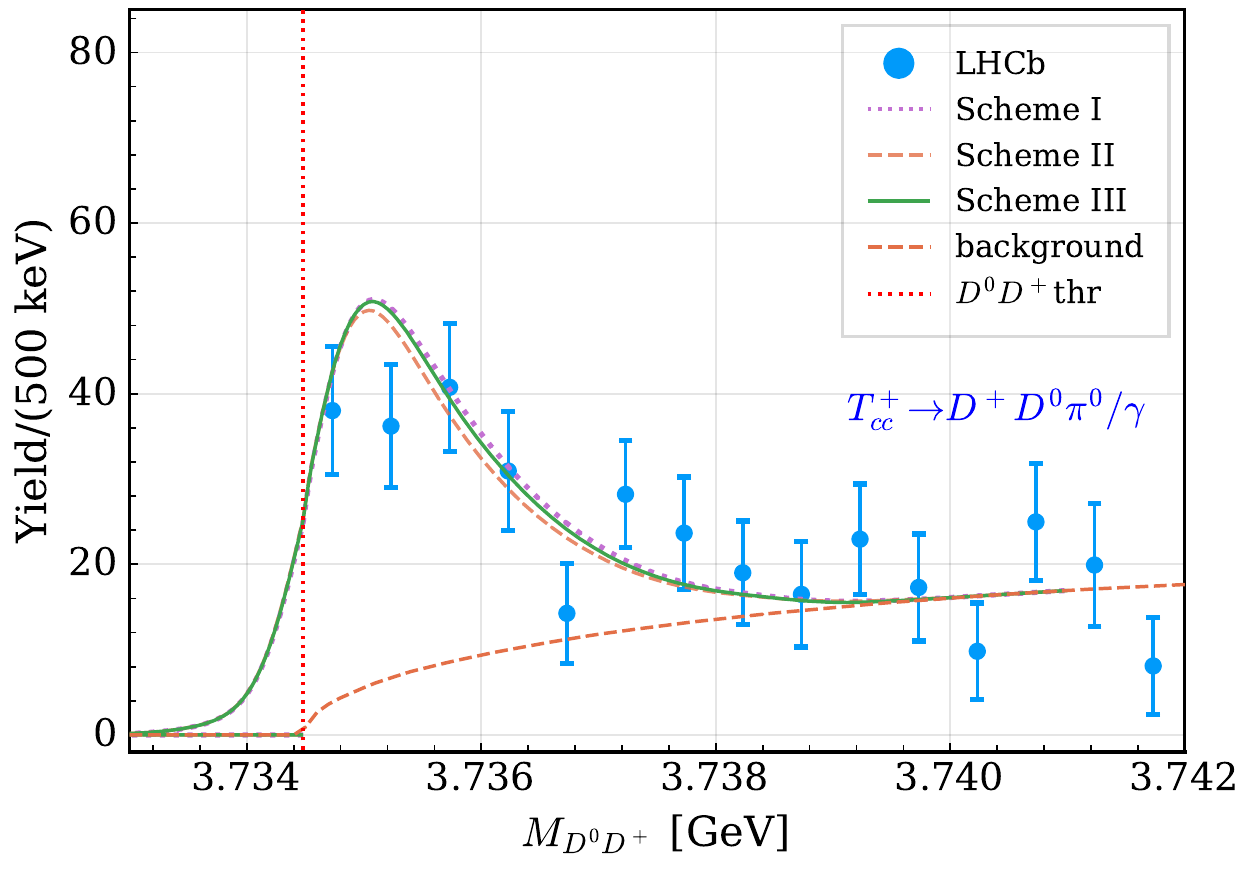}
 \caption{The same as in Fig.~\ref{fig:D0D0} but for the $D^0D^+$ invariant mass distribution of the $T_{cc}^+\to D^0D^+\pi^0/\gamma$ decays.}
 \label{fig:D0Dc}
\end{figure*}

\subsection{Heavy-quark spin symmetry partners of $\tcc$}\label{sec:partner}
According to HQSS, in a given isospin sector, the interaction of certain $D^{(*)}D^{(*)}$ pairs can be related to that of the $D^*D$ in the heavy-quark limit---see, for example, Appendix~\ref{app:su3}. In particular, 
\bea\label{eq:vhqss}
V^{I=0}(D^*D^*\to D^* D^*,1^+) & = & V^{I=0}(D^*D \to D^* D, 1^+) ,
\eea
which holds for both the contact and OPE potentials and is generally true for all interactions that preserve HQSS. The isoscalar $\tcc$ hints at the possible existence of a heavy-quark spin partner in the isoscalar $J^P=1^+$ sector. An additional $D^*D^*$ state, $T_{cc}^{*+}$ as introduced in Ref.~\cite{Albaladejo:2021vln}, can be predicted based on Eq.~\eqref{eq:vhqss}, provided that the $DD^*$-$D^*D^*$ coupled-channel effects are neglected. 

While being of little effect for the $\tcc$ itself, $D$-waves (and especially $S$-$D$ transitions) are known to play an important role for a precise
prediction of the spin partner states, since the energy region covered by the effective field theory needs to be quite large (about 150~MeV in the charm sector that results in momenta as large as 500~MeV)~\cite{Baru:2016iwj}. Renormalization of the OPE in the theory with such a large applicability domain requires an inclusion of higher, momentum-dependent contact interactions and, to fix those, 
experimental data with a nontrivial signal far beyond just a narrow near-threshold region are needed. Such studies were performed recently in the context of the $Z_b(10610)$ and $Z_b(10650)$~\cite{Wang:2018jlv,Baru:2019xnh} and the LHCb pentaquarks~\cite{Du:2021fmf}. 

Given the very limited data on the double-charm sector available at the moment, a precise prediction of the $T_{cc}^{*+}$ is not possible yet, and cannot go further than providing naive estimates based solely on identical LO contact potentials in different channels. 
Nonetheless, as an estimation, we could simply neglect the $DD^*$-$D^*D^*$ coupled-channel effects and try to test the significance played by the OPE potential based on Eq.~\eqref{eq:vhqss}. Once the coupled channels are neglected, the OPE potentials for the $D^*D^*\to D^* D^*$ transitions have no three-body cut since there is no $D^\ast\to D\pi$ vertex. It should be noted that the {\B $D^*D^*$} propagator contains a four-body cut 
due to the self-energies of two $D^*$ that
is, however, a higher-order effect lying beyond the scope of this work. Therefore, as long as only the mass of the spin partner of the $\tcc$ is concerned, it is safe to neglect the $D^*$ width.\footnote{One can also use a complex $D^*$ mass to include its width, as in Scheme I. We have explicitly checked that the inclusion of constant widths of the $D^*$'s does not affect the mass of the $T_{cc}^{*+}$ in Schemes I and II. For Scheme III, the effect of the $D^*$ width on the mass of the $T_{cc}^{*+}$ is around 30~keV, that is, well within the uncertainty quoted in Eq.~\eqref{Eq:Ebpartner}.
} 
We find that the isoscalar $D^*D^*$ amplitude with $J^P=1^+$ possesses a pole on the first Riemann sheet with the binding energy (the real part relative to the $D^*D^*$ threshold)
\bea
&\text{Scheme I: } & \quad \delta_{cc}^{*+} = -1444(61)~ \text{keV}, \nonumber\\
&\text{Scheme II: } & \quad\delta_{cc}^{*+} = -1138(50)~ \text{keV}, \label{Eq:Ebpartner}\\
&\text{Scheme III:} &\quad \delta_{cc}^{*+} = -503(40) ~\text{keV}, \nonumber
\eea
where $ \delta_{cc}^{*+} = m_{T_{cc}^{*+}}-m_c^*-m_0^*$. A large spread in the predictions for the mass of the $\tcc$ spin partner in the three schemes employed implies a possibly significant role of the OPE interaction. 

Further considerations, which might be relevant for hypothetical SU(3) siblings of the $\tcc$ and $T_{cc}^{*+}$ containing $\bar s$ antiquark(s) can be found at the end of Appendix~\ref{app:su3}.

\section{Low-energy expansion of the amplitude}
\label{sec:effective range expansion}

In this section we discuss the low-energy expansion of the scattering amplitude $D^*D\to D^*D$ and extract the corresponding parameters. 
The absolute value of the $D^*D$ scattering amplitude in the isospin basis is depicted in Fig.~\ref{fig:scattering}. 
The low-energy $S$-wave scattering parameters such as the scattering length $a_0$ and the effective range $r_0$ can be determined by scrutinizing the behavior of the scattering amplitude in the vicinity of the $\dcn$ threshold. These parameters are defined using the effective range expansion of the scattering amplitude as
\bea
T_{\dcn\to\dcn}(k) = -\frac{2\pi}{\mu_{c0}}\left({\frac{1}{a_0} + \frac12 r_0 k^2 -ik}+\mathcal{O}(k^4)\right)^{-1}.
\label{eq:reff}
\eea

It is important to notice that the finite width of the $D^*$ drives the three-momentum $k$ ill-defined in the vicinity of the two-body $D^*D$ threshold (a detailed discussion of this and related issues can be found in Refs.~\cite{Braaten:2007dw,Hanhart:2010wh}, the problem is revisited in a recent work \cite{Baru:2021ldu}). 
In order to get a deeper insight into this problem, let us start from a single-channel study with a constant contact potential $V_{\rm CT}$. This corresponds to a single-channel version of our Scheme~I. Then it is easy to find that the inverse scattering amplitude is simply
\be
T^{-1}(M) = V_{\rm CT}^{-1}+J(M),\quad J(M) = \int\frac{\D^3\vec{p}}{(2\pi)^3} G(M,p),
\label{eq:gfunction}
\ee
where $G(M,p)$ is the Green's functions of the form as defined in Eq.~\eqref{eq:twoprop}. Therefore, in this trivial example, the effective range is just $r_0\propto -\mathfrak{Re}\frac{\D J(M)}{\D M}\bigg|_{M=M_{\rm thr}+0^+}$ with $M_{\rm thr}$ 
for the corresponding two-body threshold\footnote{Note that the limit ${M\to M_{\rm thr}}$ has to be taken from above the threshold, as indicated by $0^+$, since below the threshold the analytic continuation of the momentum $k$ also contributes to $J(M)$. This conclusion survives in the presence of three-body unitarity, as shown in Appendix~\ref{app:width}, where it is demonstrated that the effective range calculated naively from below the two-body threshold diverges in the limit of an infinitely small width. }. However, a finite width of the $D^*$ significantly modifies the behavior of $J(M)$ near the two-body threshold, since the sharp cusp is smeared by the $D^*$ width (see Fig.~\ref{fig:regcons}). Therefore, the effective range 
expansion in the vicinity of the $D^*D$ threshold has a very small radius of convergence set by the nearby complex $D^* D$ branch point, namely, $k\leqslant \sqrt{\mu_{c0} \Gamma_{D^{*+}}} \approx 9$~MeV~\cite{Baru:2021ldu}; see also Appendix~\ref{app:width}.

A way to bypass this problem is to use a complex $D^*$ mass in the relation between the energy $M$ and momentum $k$,
\be
M=m_c^*-i \Gamma_c/2+m_0+\frac{k^2}{2\mu_{c0}}.\label{eq:erepoint}
\ee
Then the expansion point $k\to 0$ is now equivalent to $M=m_c^*-i \Gamma_c/2+m_0$ in the complex energy plane, which is nothing but the branch point for the two-body unitarity cut on the unphysical RS~\cite{Braaten:2007dw}. In other words, the effective range expansion is defined around the pole of the Green's function $G(M,p)$. That this holds true approximately also in the presence of three-body unitarity was demonstrated in Ref.~\cite{Hanhart:2010wh} (see Figs. 2 and 3 in the cited work). Formulated in this way, the suggested approach is straightforwardly generalized to Schemes II and III.\footnote{In these schemes RS-I and RS-II are continuously connected to each other along the three-body cut on the real axis.} 
Then, the effective range expansion with the scattering parameters
extracted in this way and collected in Table~\ref{tab:reff_2} matches very well the exact amplitude in the low-energy region of interest, especially in the vicinity of the $D^*D$ threshold, as naturally expected for a properly defined low-energy expansion. 
It is instructive to note that the scattering length extracted at the complex threshold slightly deviates from the value of the amplitude at the nominal $D^{*+}D^0$ threshold, $-({\mu_{c0}}/{(2\pi))}T_\text{thr}$, where the threshold $m_{\rm thr}=m_c^*+m_0$ is real by definition (see Table~\ref{tab:reff_2}).

\begin{figure*}[t!]
 \centering
 \includegraphics[width=0.5\textwidth]{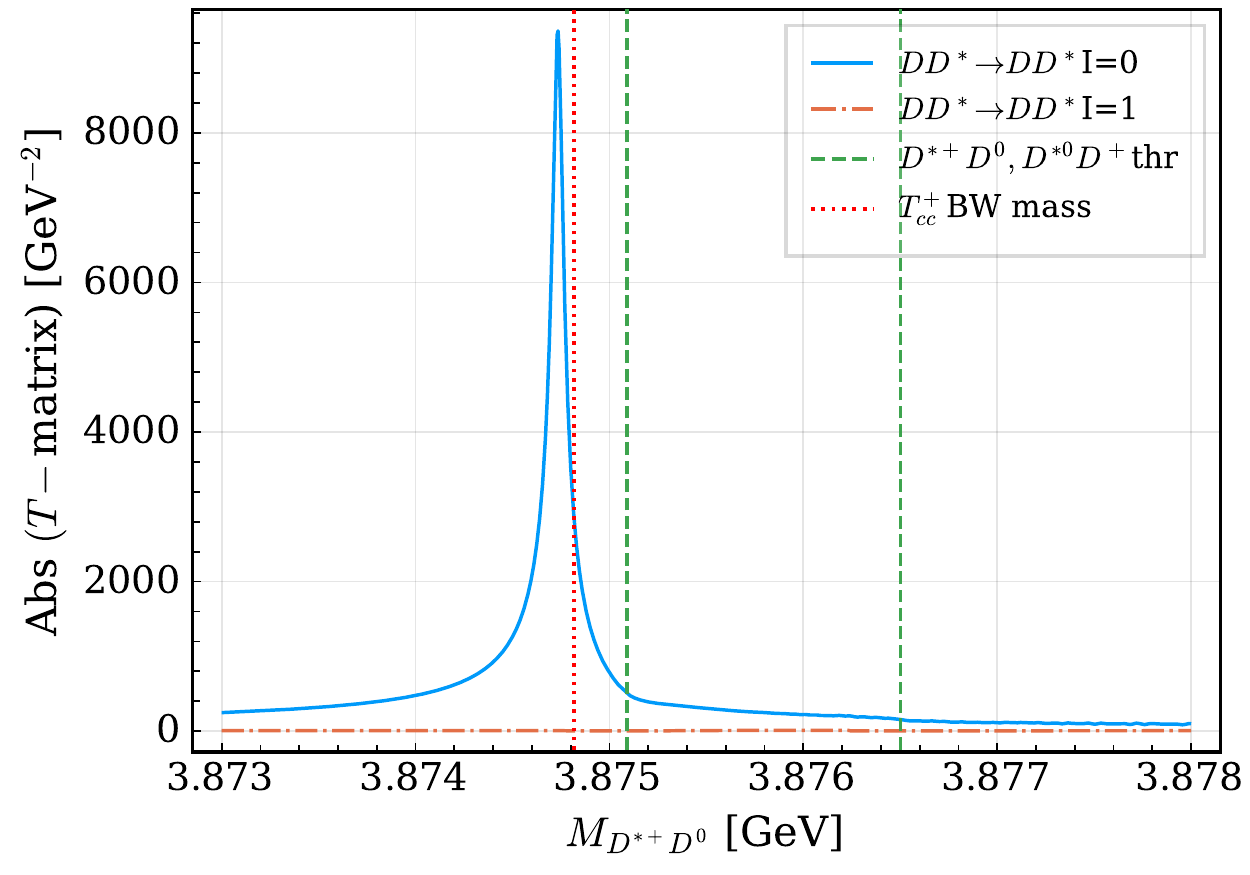}
 \caption{The absolute value of the $D^*D\to D^*D$ scattering amplitudes for the parameters of Scheme III (see Table~\ref{tab:fits}).}
 \label{fig:scattering}
\end{figure*}

\begin{figure*}[t!]
\centering
\includegraphics[width=0.5\textwidth]{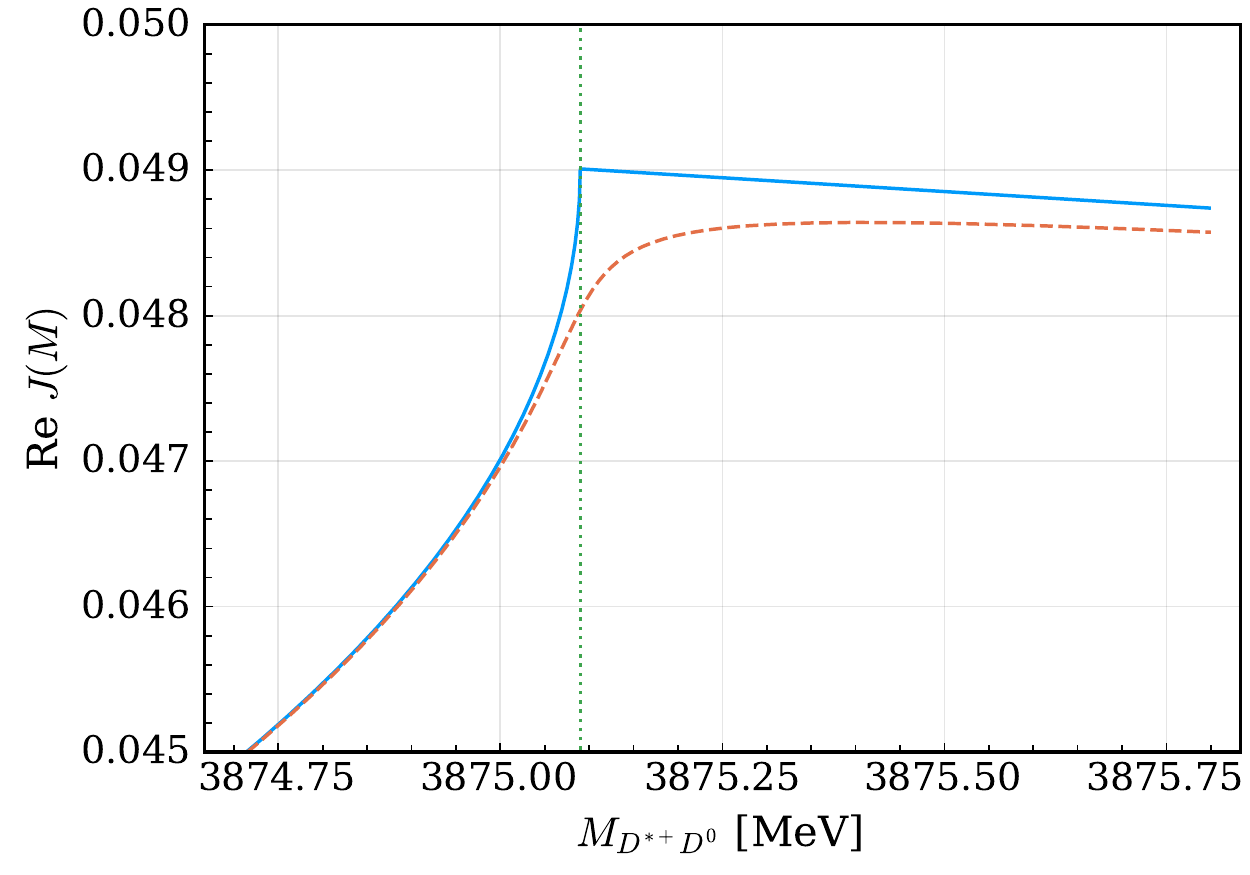}
\caption{Real part of the single-channel loop function $J(M)$ introduced in Eq.~\eqref{eq:gfunction} for a zero (blue solid line) or finite (red dashed line) constant $D^{*+}$ width. The vertical dotted line marks the $\dcn$ threshold.}
\label{fig:regcons}
\end{figure*}

\begin{table*}
 \begin{ruledtabular}
 \caption{The $S$-wave scattering length $a_0$ and effective range $r_0$ parameters defined in Eq.~\eqref{eq:reff} extracted as explained in the text (see the discussion around Eq.~\eqref{eq:erepoint}). The compositeness $\bar X_A$ is obtained using Eq.~\eqref{eq:XA} and the compositeness $X_1$ and $X_2$ are calculated using Eq.~\eqref{eq:Xis}. Because of the finite widths of the $D^*$s, the results for the $X_i$'s become complex, though their imaginary parts are negligible and, therefore, only the real parts are given
in the Table. For each value, the error in the first line is statistical propagated from fitting to the LHCb data; the error in the second line is systematic from model uncertainty. It is estimated by varying the cutoff parameter $\Lambda$ in the interval of $[0.3,1.2]$~GeV and taken as the largest deviation from the central value. The central value is evaluated for $\Lambda=0.5$~GeV. }
 {\fontsize{9}{10}
 \begin{tabular}{l| c|cc|c|cc}
 & $-\dfrac{\mu_{c0}}{2\pi}T_\text{thr}$ [fm] & $a_0$ [fm] & $r_0$ [fm]& $\bar X_A$ & $X_1$ & $X_2$ \\
 \hline
 I & $\bigg(\parbox{0mm}{\begin{align}
 -7.38&^{+0.46}_{-0.57} \notag \\[-2mm]
 & \scalebox{0.8}{$\pm0.36$} \notag
 \end{align}}\bigg)\!\! +\! i\bigg(\parbox{0mm}{\begin{align}
 1.96&^{+0.34}_{-0.57} \notag \\[-2mm]
 & \scalebox{0.8}{$\pm0.18$} \notag
 \end{align}}\bigg)$ & 
 $\bigg(\parbox{0mm}{\begin{align}
 -6.31&^{+0.36}_{-0.45} \notag \\[-2mm]
 & \scalebox{0.8}{$\pm0.27$} \notag
 \end{align}}\bigg)\!\! +\! i\bigg(\parbox{0mm}{\begin{align}
 0.05&^{+0.01}_{-0.01} \notag \\[-2mm]
 & \scalebox{0.8}{$\pm0.00$} \notag
 \end{align}}\bigg)$
 & \parbox{0mm}{\begin{align}
 -2.78&\pm0.01 \notag \\[-2mm]
 & \pm0.66 \notag
 \end{align}} & 
 \parbox{0mm}{\begin{align}
 0.87&\pm0.01 \notag \\[-2mm]
 & \pm0.07 \notag
 \end{align}} & 
 \parbox{0mm}{\begin{align}
 0.71&\pm0.01 \notag \\[-2mm]
 & \pm0.02 \notag
 \end{align}} & \parbox{0mm}{\begin{align}
 0.29&\pm0.01 \notag \\[-2mm]
 & \pm0.02 \notag
 \end{align}}\\[-5mm]
 II & $\bigg(\parbox{0mm}{\begin{align}
 -8.00&^{+0.49}_{-0.68} \notag \\[-2mm]
 & \scalebox{0.8}{$\pm0.35$} \notag
 \end{align}}\bigg)$\!\! +\! $i\bigg(\parbox{0mm}{\begin{align}
 1.88&^{+0.36}_{-0.24} \notag \\[-2mm]
 & \scalebox{0.8}{$\pm0.18$} \notag
 \end{align}}\bigg)$
 & $\bigg(\parbox{0mm}{\begin{align}
 -6.64&^{+0.36}_{-0.50} \notag \\[-2mm]
 & \scalebox{0.8}{$\pm0.27$} \notag
 \end{align}}\bigg)\!\! -\! i\bigg(\parbox{0mm}{\begin{align}
 0.10&^{+0.01}_{-0.02} \notag \\[-2mm]
 & \scalebox{0.8}{$\pm0.01$} \notag
 \end{align}}\bigg)$ & 
 \parbox{0mm}{\begin{align}
 -2.80&\pm0.01 \notag \\[-2mm]
 & \pm0.59 \notag
 \end{align}}
 & \parbox{0mm}{\begin{align}
 0.88&\pm0.01 \notag \\[-2mm]
 & \pm0.06 \notag
 \end{align}} 
 & \parbox{0mm}{\begin{align}
 0.71&\pm0.01 \notag \\[-2mm]
 & \pm0.02 \notag
 \end{align}}
 & \parbox{0mm}{\begin{align}
 0.29&\pm0.01 \notag \\[-2mm]
 & \pm0.02 \notag
 \end{align}} \\[-5mm]
 III & $\bigg(\parbox{0mm}{\begin{align}
 -7.76&^{+0.45}_{-0.53} \notag \\[-2mm]
 & \scalebox{0.8}{$\pm0.32$} \notag
 \end{align}}\bigg)$\!\! +\! $i\bigg(\parbox{0mm}{\begin{align}
 2.44&^{+0.38}_{-0.29} \notag \\[-2mm]
 & \scalebox{0.8}{$\pm0.18$} \notag
 \end{align}}\bigg)$
 & $\bigg(\parbox{0mm}{\begin{align}
 -6.72&^{+0.36}_{-0.45} \notag \\[-2mm]
 & \scalebox{0.8}{$\pm0.27$} \notag
 \end{align}}\bigg)\!\! -\! i\bigg(\parbox{0mm}{\begin{align}
 0.10&^{+0.03}_{-0.03} \notag \\[-2mm]
 & \scalebox{0.8}{$\pm0.03$} \notag
 \end{align}}\bigg)$
 & \parbox{0mm}{\begin{align}
 -2.40&\pm0.01 \notag \\[-2mm]
 & \pm0.85 \notag
 \end{align}}
 & \parbox{0mm}{\begin{align}
 0.84&\pm0.01 \notag \\[-2mm]
 & \pm0.06 \notag
 \end{align}}
 & \parbox{0mm}{\begin{align}
 0.73&\pm0.01 \notag \\[-2mm]
 & \pm0.11 \notag
 \end{align}} 
 & \parbox{0mm}{\begin{align}
 0.27&\pm0.01 \notag \\[-2mm]
 & \pm0.02 \notag
 \end{align}} \\
 \end{tabular}
 }
 \end{ruledtabular}
 \label{tab:reff_2}
 \end{table*}

As shown in Ref.~\cite{Baru:2021ldu}, 
in the approach involving $D^{*+}D^0$ and $D^{*0}D^+$ coupled channels, the largest contribution to the effective range for the $T_{cc}^+$ originates from isospin breaking (IB) related to the $D^{(*)}$-meson mass differences, that is, from the coupling of $D^{*+}D^0$ to the slightly higher $D^{*0}D^+$ channel. This gives 
 \be\label{Eq:rIV}
 \Delta r_{\rm IB} = - \sqrt{\frac{\mu_{c0}}{2\mu_{0c}^2(m_0^*+m_c-m_c^*-m_0)}} \approx -3.78~\mbox{fm}.
 \ee
By comparing this result with the effective range for Schemes I and II in Table~\ref{tab:reff_2}, one finds, in line with Ref.~\cite{Baru:2021ldu}, that the residual finite range correction is
$\simeq 1$ fm for the cutoff $\Lambda= 0.5$ GeV, and it may be reduced to 0.5 fm if the cutoff is increased to 1 GeV. On the other hand, the comparison of the effective range for Schemes III and I/II shows the difference $\simeq 0.4$ fm, which is the effect from the OPE.

We note that the finite range corrections to the effective range $r_0$ are proportional to ${\Lambda}^{-1}$~\cite{Baru:2021ldu}, where $\Lambda$ can be regarded as a scale corresponding to heavier-meson exchange contributions not included here explicitly. It is instructive to estimate the model uncertainty of the ERE parameters 
due to the variation of the cutoff parameter $\Lambda$. We, therefore, vary it in a rather wide range $[0.3,1.2]$~GeV, consistent with the effective field theory built in this work and used to fit the LHCb data.
The systematic model uncertainty of each quantity is then associated with the largest deviation from its central value calculated for $\Lambda=0.5$~GeV.
These model uncertainties are also collected in Table~\ref{tab:reff_2}, as shown in the second line for each scheme.
From this table one can see that the model uncertainty for the scattering length $a_0$ is comparable or less than the statistical one. On the contrary, for the effective range $r_0$ the model uncertainty dominates, which 
should not come as a surprise given that the effective theory is constructed to LO.
To improve on that, the next-to-leading ${\cal O}(k^2)$ contact potential needs to be invoked to absorb the residual $\Lambda$-dependence of the effective range which, however, would not be accurately fixed by the limited data currently available. Meanwhile, we still find
that in our 
effective field theory approach 
the effective range is constrained better than in the LHCb analysis~\cite{LHCb:2021auc}.

\section{Composite or compact?}
\label{sec:XA}

With the parameters of the low-energy expansion of the amplitude reliably extracted from the fit to the experimental data, we are in a position to address the question of the nature of the $\tcc$ state. Namely, whether or not our original assumption on the $\tcc$ as a $D^*D$ molecule is supported by a formal calculation of its compositeness parameter 
$\bar X_A$ constructed from the scattering length and effective range as~\cite{Weinberg:1965zz,Matuschek:2020gqe}
\bea
\bar X_A = \left({1+2\left| \frac{r_0^\prime}{\mathfrak{Re}\,a_0} \right| }\right)^{-1/2},
\label{eq:XA}
\eea
where {$r_0^\prime = r_0 - \Delta r_{\rm IB}$} is the effective range in the 
$D^{*+}D^0$ channel
after 
the coupled-channel isospin-violation corrections defined in Eq.~\eqref{Eq:rIV} were subtracted
\cite{Baru:2021ldu}.

Then $\bar X_A\simeq 1$ would correspond to a composite state formed by the $D$ and $D^*$ while $\bar X_A\ll 1$ would imply a compact state. A similar formulation was used by Weinberg in its original paper~\cite{Weinberg:1965zz} where the measured low-energy parameters (the scattering length and effective range) were formally demonstrated to be consistent with the deuteron being a compound rather than an elementary state.\footnote{For detailed discussions of the positive effective range case, as it occurs for the deuteron, see Refs.~\cite{Matuschek:2020gqe,Li:2021cue}. } 

The compositeness $\bar X_A$ estimated for each scheme with the help of Eq.~(\ref{eq:XA}) and the values of the $a_0$ and $r_0$ listed in Table~\ref{tab:reff_2} is given in the fifth column of the same table. One can see that indeed $\bar{X}_A\simeq 1$, so that the assumption of the molecular nature of the $\tcc$ is fully justified. Furthermore, as an independent additional check, we evaluate the compositeness of this resonance in each coupled channel involved individually~\cite{Hyodo:2011qc,Aceti:2012dd,Sekihara:2014kya,Guo:2015daa},
\be
X_i = g_i^2\frac{\D J_i(M_\text{pole})}{\D M^2_\text{pole}},
\label{eq:Xis}
\ee
where $i=\dcn$, $\dnc$ and $g_i$'s are the couplings of the $\tcc$ to the $i$th channel listed in Table~\ref{tab:poles}. 
It is instructive to notice that the values of $X_1$ and $X_2$ given in Table~\ref{tab:reff_2} sum exactly to unity thus ensuring the $\tcc$ to be indeed a molecular state with the relevant set of coupled channels saturated by the $\dcn$ and $\dnc$.

\section{Discussion}
\label{sec:disc}

In this work we employed a coupled-channel approach based on a nonrelativistic effective field theory to analyze the experimental data on the charged double-charm meson $\tcc$ recently discovered by the LHCb Collaboration. 
The effective potential at LO 
includes one isoscalar momentum-independent contact interaction and the OPE. 
The data are found to be fully consistent with the $\tcc$ being a weakly bound $D^*D$ molecule, with the corresponding pole lying just below the $D^{*+}D^0$ threshold. The found pole can be associated with a bound state with respect to the two-body channels, with the reservation that its nomenclature formally depends on the way the three-body dynamics is introduced: it appears slightly shifted from the real axis if constant $D^*$'s widths are included effectively or should be regarded as a pole on an unphysical RS if dynamical $D^*$'s widths are considered and the three-body cuts are introduced. 

A self-consistent treatment of the three-body dynamics is our special concern, given the lessons previously learnt from the studies of another near-threshold charmoniumlike state: the $X(3872)$. In particular, we emphasize that the constraints of three-body unitarity can only be fulfilled if both dynamical ingredients, momentum-dependent $D^*$'s widths due to the $D^*\to D\pi$ decay channels and the OPE between the $D$ and $D^*$ mesons ($\tcc$ constituents) are included simultaneously. 
The inclusion of only one effect from those mentioned above, as in Ref.~\cite{LHCb:2021auc}, results in a violation of three-body unitarity and leads to unreliable results, especially for the precise determination of the $\tcc$ pole position in the complex energy plane. On the other hand, a simultaneous neglect of both of these three-body effects, and restricting oneself to the static OPE together with the use of constant widths for the $D^*$'s
produces a severe 
overestimation of the $\tcc$ width, evaluated as twice the imaginary part of the $\tcc$ pole.\footnote{Needless to say that a naive Breit-Wigner fit to the data~\cite{LHCb:2021vvq} returns the $\tcc$ width far beyond the theoretically acceptable values.} For this reason the appropriate treatment of the OPE appears to be an important 
prerequisite of any precise 
approach to the $\tcc$ and, unless the opposite is proven, other narrow near-threshold states wherever the pion exchange introduces nearby three-body branch points.

The low energy expansion of the amplitude is performed along the lines of the recent work~\cite{Baru:2021ldu} and the scattering length and effective range are extracted. 
It is observed that the effective range takes moderate negative values which are however very well understood within the molecular picture, as they come from the $\dnc$ channel, coupled to $\dcn$, that has a slightly higher threshold than that of the $\dcn$ due to isospin symmetry breaking. Once this contribution is subtracted, the residual effective range appears to be small and positive in line with the expected size of the finite-range corrections. This residual effective-range contribution is directly related to the coupling of the $\tcc$
to its constituents, while the isospin 
breaking effects in this coupling can be safely neglected.

Equipped with the low-energy parameters extracted from data, we have evaluated the compositeness of the $\tcc$ state using two methods. As expected, both provide similar results, and we have found that the data are fully consistent with the $\tcc$ as a shallow $D^*D$ molecular state with the compositeness close to unity. 

Finally, based on HQSS, we have predicted the existence of an isospin scalar $D^*D^*$ molecule with $J^P=1^+$, under the assumption that the $DD^*$-$D^*D^*$ coupled-channel effects can be neglected.
We found in particular that the inclusion of the OPE has a visible impact on its binding energy, moving the state towards the threshold.
However, a precise prediction for the spin partner of the $\tcc$ is unlikely possible until the OPE potential and $DD^*$-$D^*D^*$ coupled-channel effects are properly taken into account, which is not possible yet given the limited data currently available on the double charm sector.

\medskip

\begin{acknowledgments}
We would like to thank Marek Karliner for useful discussions regarding the pion exchange and Vanya Belyaev for providing information on the energy resolution function, and Alexander Bondar for reading the manuscript and valuable comments. 
This work is supported in part by the Spanish Ministry of Science and Innovation (MICINN) (Project PID2020-112777GB-I00), by the EU Horizon
2020 research and innovation programme, STRONG-2020 project, under grant agreement No.~824093, by Generalitat
Valenciana under contract PROMETEO/2020/023, by the National Natural Science Foundation of China (NSFC) and the
Deutsche Forschungsgemeinschaft (DFG) through the funds provided to the Sino-German Collaborative Research Center TRR110 ``Symmetries and the Emergence of Structure in QCD'' (NSFC Grant No. 12070131001, DFG Project-ID 196253076),
by the NSFC under Grants No.~11835015, No.~12047503, No.~11961141012, and No.~12035007, and by the Chinese Academy of Sciences under Grants
No. QYZDB-SSW-SYS013, No. XDB34030000, No.~XDPB15 and No.~2020VMA0024.
The work of A.N. is supported by the Ministry of Science and Education of the Russian Federation under grant 14.W03.31.0026. The work of Q.W. is also supported by Guangdong Major Project of Basic and Applied Basic Research under Grant No.~2020B0301030008, by the
Science and Technology Program of Guangzhou under Grant No.~2019050001, and by
Guangdong Provincial funding under Grant No.~2019QN01X172.
\end{acknowledgments}

\begin{appendix}

\section{Contact potentials in the isospin and SU(3) limit}\label{app:su3}

The contact potentials for the different isospin/spin-parity $D^{(*)}D^{(*)}\to D^{(*)}D^{(*)}$ $S$-wave channels derived from the Lagrangian of Eq.~(\ref{eq:LHQSS}) read
\begin{align}
V_{\rm CT}^{I=0}(D^*D\to D^*D;1^+)&=-2(D_{01}-3D_{11}),\no
V_{\rm CT}^{I=0}(D^*D^*\to D^*D^*;1^+)&=-2(D_{01}-3D_{11}),\label{eq:A1}\\
V_{\rm CT}^{I=0}(D^*D\to D^*D^*;1^+)&=-D_{00}+D_{01}+3D_{10}-3D_{11}, \nonumber
\end{align}
with $V_{\rm CT}^{I=0}\left(D^{(*)}D^{(*)}\to D^{(*)}D^{(*)};0^+\right)=V_{\rm CT}^{I=0}(D^*D^*\to D^*D^*;2^+)=0$ as per Bose statistics, and 
\begin{align}
V_{\rm CT}^{I=1}(DD\to DD;0^+)&=\frac12(D_{00}+3D_{01}+D_{10}+3D_{11}), \no
V_{\rm CT}^{I=1}(D^*D^*\to D^*D^*;0^+)&=-\frac{1}2(D_{00}-5D_{01}+D_{10}-5D_{11}),\no
V_{\rm CT}^{I=1}(DD\to D^*D^*;0^+)&=-\frac{\sqrt3}{2}(D_{00}-D_{01}+D_{10}-D_{11}),\\
V_{\rm CT}^{I=1}(D^*D\to D^*D;1^+)&=D_{00}+D_{01}+D_{10}+D_{11},\no
V_{\rm CT}^{I=1}(D^*D^*\to D^*D^*;2^+)&=D_{00}+D_{01}+D_{10}+D_{11},\nonumber
\end{align}
with $V_{\rm CT}^{I=1}(D^* D\to D^*D^*;1^+)=V_{\rm CT}^{I=1}(D^*D^*\to D^*D^*;1^+)=0$ (Bose statistics). Here we used that the isoscalar (isovector) wave function for two identical particles with isospin $I=1/2$ are antisymmetric (symmetric), respectively. 

There are in total four independent LECs, which can be assigned to configurations for the light-quark subsystem in the $D^{(*)}D^{(*)}$ wave function with $I=0,1$ and degrees of freedom coupled to the spin $s_\ell=0,1$~\cite{Nieves:2012tt, Hidalgo-Duque:2012rqv,Guo:2013sya, Guo:2017jvc}. Denoting these new LECs by $D^{I=0,1}_{s_\ell=0,1}$, one readily finds that
\begin{eqnarray}
D^{I=0}_{s_\ell=1} &=& 3 D_{10}+3 D_{11}-D_{00}-D_{01}, \quad D^{I=1}_{s_\ell=1} = D_{00}+D_{01}+D_{10}+D_{11}, \nonumber \\[-2mm]
\\[-2mm]
D^{I=0}_{s_\ell=0} &=& D_{00} -3 D_{10}-3 D_{01}+9 D_{11}, \quad D^{I=1}_{s_\ell=0} = 3 D_{01}+3 D_{11}-D_{00}-D_{10}.\nonumber
\end{eqnarray}

The extension of the contact potentials quoted above to the light quark flavor SU(3) sector is straightforward and amounts to replacing the Pauli matrices $\tau$ by the Gell-Mann ones $\lambda$ and including the antistrange-charmed mesons in the superfield $H$. Then the light-quark subsystem in the $HH$ system can be decomposed into SU(3) irreducible representations (irreps) as $\bar3\otimes\bar3=3\oplus\bar6$. 
Hence, there are still only four independent LECs, $D^{R=3,\bar 6}_{s_\ell=0,1}$, which correspond to the $3$ and $\bar 6$ flavor configurations combined with $s_\ell=0,1$. The $3$ irrep of SU(3) contains one isospin singlet ($I=0)$ and one doublet ($I=1/2$) and the $\bar6$ irrep contains one isospin triplet ($I=1$), one doublet and one singlet. It trivially follows that 
\begin{equation}
D^{R=\bar 6}_{s_\ell=0,1}= D^{I=1}_{s_\ell=0,1}\ , \quad D^{R=3}_{s_\ell=0,1}= D^{I=0}_{s_\ell=0,1}\, , 
\end{equation}
since the isoscalar state in the $3$ irrep of SU(3) does not involve the charm-strange mesons while that in the $\bar 6$ is constructed out of the $D_s^{(*)}D_s^{(*)}$ pair~\cite{Garcia-Recio:2010dfh}. The identification of the four LECs $D^{R=3,\bar 6}_{s_\ell=0,1}$ allows to describe the contact interaction of any $D^{(*)}_{(s)} D^{(*)}_{(s)} \to D^{(*)}_{(s)} D^{(*)}_{(s)}$ transition, for any isospin ($I=0,1/2,1$) and $J^P=0^+,1^+,2^+$, in terms of the LECs which appear in the Lagrangian of Eq.~\eqref{eq:LHQSS}, as was done in Ref.~\cite{Hidalgo-Duque:2012rqv} for the $H \bar H$ case.\footnote{Equivalently, if we denote by $D^\prime_{00,01,10,00}$ the constants of a Lagrangian as that of Eq.~\eqref{eq:LHQSS}, but where the SU(2) Pauli matrices have been replaced by the SU(3) Gell-Mann ones, these new LECs will be given by $D^\prime_{00}= D_{00} - D_{10}/3$, $D^\prime_{01}= D_{01} - D_{11}/3$, $D^\prime_{10}= D_{10}$ and $D^\prime_{11}= D_{11}$.} For example, 
\begin{align}
 V_{\rm CT}(D_s^*D_s^*\to D_s^*D_s^*;2^+) &=D^{R=\bar 6}_{s_\ell=1}=V_{\rm CT}^{I=1}(D^*D^*\to D^*D^*;2^+), \nonumber\\
 V_{\rm CT}(D_s^*D_s^*\to D_s^*D_s^*;0^+) &= \frac14\left(3 D^{R=\bar 6}_{s_\ell=0}+D^{R=\bar 6}_{s_\ell=1}\right) =V_{\rm CT}^{I=1}(D^*D^*\to D^*D^*;0^+).
\end{align}

{
The $\tcc$ would be placed in the $3$ irrep of SU(3) with $J^P=1^+$. We see in Eq.~\eqref{eq:A1} that $V_{\rm CT}^{I=0}(D^*D\to D^*D;1^+)= V_{\rm CT}^{I=0}(D^*D^*\to D^*D^*;1^+)=-2(D_{01}-3D_{11}) = (D^{R=3}_{s_\ell=0}+D^{R=3}_{s_\ell=1})/2$, and hence from the isoscalar factors compiled in Ref.~\cite{Garcia-Recio:2010dfh}, we would find the same contact interactions for the $[D^*D_s]_A\to [D^*D_s]_A$ and $[D^*D^*_s]_A\to [D^*D^*_s]_A$ transitions in the $cc\bar s$ sector. Here the subscript $A$ stands for normalized anti-symmetric states $[D^*D_s]_A= (D^*D_s-D^*_sD)/\sqrt{2}$ and $[D^*D^*_s]_A= (D^*D^*_s-D^*_sD^*)/\sqrt{2}$. As a consequence, there might exist strange partners with $J^P=1^+$ of the $\tcc$ in the 
$D^*D_s$ and $D^*D^*_s$ channels with masses around the corresponding thresholds, as a result of the common contact interaction $(D^{R=3}_{s_\ell=0}+D^{R=3}_{s_\ell=1})/2$. These predictions are obviously subject to SU(3) symmetry breaking corrections. }

\section{The effect of a finite width on the effective range}
\label{app:width}

In this appendix we employ a simple coupled-channel model to investigate the influence of the three-body dynamics on the effective range for a near-threshold resonance $X$. We follow the lines of Ref.~\cite{Hanhart:2010wh} and consider a Fock space consisting of three states: a compact seed, labelled as $X_0$, an $ab$ pair in their relative $S$-wave, and the $a[cd]$ state, where the particles $c$ and $d$ are the products of the decay $b\to c+d$ which proceeds in a partial wave defined by the angular momentum $l$ (the discussion around Eq.~\eqref{eq:gfunction} corresponds to a $P$-wave decay with $l=1$). Thus the wave function of the resonance $X$ can be written as
\be
|X\rangle=
\left(
\begin{array}{c}
C|X_0\rangle\\
\chi(\vep)|ab\rangle\\
\varphi(\vep,\veq)|a[cd]\rangle
\end{array}
\right),
\label{psi}
\ee
where $\vep$ and $\veq$ are the center-of-mass momenta in the $ab$ and $cd$ subsystems, respectively. The two- and three-body thresholds, $M_{\rm thr}^{(2)}$ and $M_{\rm thr}^{(3)}$, respectively, are split by
\be
E_R=M_{\rm thr}^{(2)}-M_{\rm thr}^{(3)}>0,
\ee
and the width of the unstable constituent $b$ is
\be
\varGamma_R=g_l E_R^{l+1/2},
\label{gldef}
\ee
where $g_l$ is a coupling constant which governs the decay $b\to c+d$. It also proves convenient to define a dimensionless ratio
\be
\lambda=\frac{\varGamma_R}{2E_R},
\ee
which in what follows will be assumed to be small, $\lambda\ll 1$, corresponding to the case of a narrow constituent.

The system of coupled-channel Lippmann--Schwinger equations for the state (\ref{psi}) was solved in Ref.~\cite{Hanhart:2010wh} for the interaction between channels exhausted by the transitions $X_0\leftrightarrow ab$ and $ab\leftrightarrow acd$
described by momentum-dependent form factors whose explicit form is not important. 
As a result, the line shapes in the three-body final state $acd$ were obtained and compared with the predictions of simple prescriptions found in the literature. The scattering amplitude in the $S$-wave two-body channel was found to be
\be
T(a+b\to a+b)=\frac{\pi}{\mu}\frac{g}{E-E_X+G_X(E)},
\ee
where the self-energy $G_X(E)$ of the resonance $X$ can be written as 
\be
\mathfrak{Re}\left(G_X(E)\right)=\frac12g\kappa_{\rm eff}(E),
\quad 
\mathfrak{Im}\left(G_X(E)\right)=\frac12gk_{\rm eff}(E),
\label{kkapdef}
\ee
with $g$ being an effective coupling constant. In the limit $\Gamma_R\to 0$ the quantities $k_{\rm eff}$ and $\kappa_{\rm eff}$ turn to the usual two-body momentum $k=\sqrt{2\mu E}$ above the two-body threshold and its analytical continuation below the threshold, respectively. Here we choose the energy $E$ to be counted from the two-body threshold $M_{\rm thr}^{(2)}$ (note also that it is counted from the lower-lying three-body threshold in Ref.~\cite{Hanhart:2010wh}, so an energy shift by $E_R$ is required to arrive from the formulas derived in Ref.~\cite{Hanhart:2010wh} to those quoted below). 

Since here we are interested in the contribution to the effective range $r_0$ coming from the width of the unstable constituent $b$, it is sufficient to consider only the function $\kappa_{\rm eff}(E)$ introduced in Eq.~(\ref{kkapdef}). Then, according to the findings of Ref.~\cite{Hanhart:2010wh},
\be
\kappa_{\rm eff}(E)=\kappa_1(E)+\kappa_2(E)-\kappa_1(E_X)-\kappa_2(E_X),
\label{kapeff}
\ee
where 
\be
\kappa_1(E)=-\frac{1}{\pi\mu}\int^{\infty}_0p^2\D p
\frac{E-\frac{p^2}{2\mu}}{(E-\frac{p^2}{2\mu})^2+
\frac{g_l^2}{4}(E+E_R-\frac{p^2}{2\mu})^{2l+1}}
\label{kap1eff}
\ee
and
\be
\kappa_2(E)=-\frac{g_l}{2\pi\mu}\int^{\infty}_{\sqrt{2\mu (E_R+E)}}
p^2\D p\frac{(\frac{p^2}{2\mu }-E_R-E)^{(2l+1)/2}}{(E-\frac{p^2}{2\mu })^2+
\frac{g_l^2}{4}(E+E_R-\frac{p^2}{2\mu})^{2l+1}}.
\label{kappa2full}
\ee

Notice that $\kappa_2(E)$ is suppressed as compared to $\kappa_1(E)$ by a factor
$g_l\propto{\lambda}$, which is small by assumption, thus in what follows we use that
\be
\kappa_{\rm eff}(E)\approx\kappa_1(E)-\kappa_1(E_X),
\ee
where the last, constant term does not contribute to the effective range evaluated as a derivative in the energy from $\kappa_{\rm eff}(E)$.

Then, using the explicit expression for $\kappa_1(E)$ from Eq.~(\ref{kap1eff}), it is easy to arrive at the contribution to the effective range
\be
\Delta r_0=-\frac{1}{\mu}\frac{\partial \kappa_{\rm eff}(E)}{\partial E}{ \bigg|_{E=0^-}}=-\frac{I(\lambda)}{\sqrt{2\mu E_R}},
\ee
where the dimensionless single-parameter function $I(\lambda)$ is
\be
I(\lambda)=\frac{1}{\pi}\int_0^\infty \D x\frac{\sqrt{x}}{x^2+\lambda^2(1-x)^{2l+1}}\mathop{\approx}_{\lambda\to 0} 
\frac{1}{\pi}\int_0^\infty \D x\frac{\sqrt{x}}{x^2+\lambda^2}=
\frac{1}{\sqrt{2\lambda}}.
\ee

Collecting all results together we finally arrive at
\be
\Delta r_0\approx -\frac{1}{\sqrt{2\mu \Gamma_R}},
\label{Dr0}
\ee
that provides an uncontrollably large negative contribution to the effective range in the limit $\Gamma_R\to 0$. 

Interpretation of the result obtained is straightforward: since the three-body threshold lies below the two-body one, for any finite width $\Gamma_R$, the self-energy $G_X(E)$ analytically continues below $M_{\rm thr}^{(2)}$ and, therefore, for $\Gamma_R\to 0$ the contribution described by Eq.~(\ref{Dr0}) turns to $-\mathfrak{Re}\frac{\D J(M)}{\D M}\bigg|_{M=M_{\rm thr}+0^-}$ which is indeed infinite in this limit---see Fig.~\ref{fig:regcons}.

\end{appendix}

\bibliography{Tcc_refs.bib}

\end{document}